\documentclass[aps,prl,preprintnumbers,showpacs,nofootinbib]{revtex4}
\usepackage{graphicx}% Include figure files
\usepackage{dcolumn}% Align table columns on decimal point
\usepackage{bm}% bold math
\begin{document}
\newcommand{\newc}{\newcommand}
\newc{\ra}{\rightarrow}
\newc{\lra}{\leftrightarrow}
\newc{\beq}{\begin{equation}}
\newc{\eeq}{\end{equation}}
\newc{\barr}{\begin{eqnarray}}
\newc{\earr}{\end{eqnarray}}
%%%%%%%%%%%%%%%%%%%%%%%%%%%%%%%%%%%%%%%%%%%
\newcommand{\Od}{{\cal O}}
\newcommand{\lsim}   {\mathrel{\mathop{\kern 0pt \rlap
  {\raise.2ex\hbox{$<$}}}
  \lower.9ex\hbox{\kern-.190em $\sim$}}}
\newcommand{\gsim}   {\mathrel{\mathop{\kern 0pt \rlap
  {\raise.2ex\hbox{$>$}}}
  \lower.9ex\hbox{\kern-.190em $\sim$}}}
%\preprint{APS/123-QED}

\title{DIRECT SUSY DARK MATTER DETECTION-\\
CONSTRAINTS ON THE SPIN CROSS SECTION
 }

\author{ J. D. Vergados\thanks{Vergados@cc.uoi.gr}}
\affiliation{
{\it Theoretical Physics Division, University of Ioannina,
Ioannina, Gr 451 10, Greece}}
%\affiliation{$^{(2)}${\it Theoretical Physics Division, University
%of Ioannina, Ioannina, Gr 451 10, Greece.}}
% \maketitle
\begin{abstract}
The recent WMAP data have confirmed that exotic dark matter
together with the vacuum energy (cosmological constant) dominate
in the flat Universe. Thus the direct dark matter detection,
consisting of detecting the recoiling nucleus, is central to
particle physics and cosmology. Supersymmetry provides a natural
dark matter candidate, the lightest supersymmetric particle (LSP).
The relevant cross sections arise out of two mechanisms: i) The
coherent mode, due to the scalar interaction and ii) The spin
contribution arising from the axial current. In this paper we will
focus on the spin contribution, which maybe important, especially
for light targets.

\end{abstract}

 %%%%%%%%%%%%%%%%%%%%%%%%%%%%%%%%%%%%%%%%%%%%%%%%%%%%%%%%%%%%%%%%%%%%
\date{\today}
%%%%%%%%%%%%%%%%%%%%%%%%%%%%%%%%%%%%%%%%%%%%%%%%%%%%%%%%%%%%%%%%%%%%%
\maketitle
\section{Introduction}
The combined MAXIMA-1 \cite{MAXIMA-1}, BOOMERANG \cite{BOOMERANG},
DASI \cite{DASI}, COBE/DMR Cosmic Microwave Background (CMB)
observations \cite{COBE}, the recent WMAP data \cite{SPERGEL} and
SDSS
%(Sloan Digital Sky Survey) results
 \cite{SDSS} imply that the
Universe is flat \cite{flat01} and
 and that most of the matter in
the Universe is dark, i.e. exotic.
  $$ \Omega_b=0.044\pm 0.04,
\Omega_m=0.27\pm 0.04,  \Omega_{\Lambda}=0.69\pm0.08$$
 for baryonic matter , cold dark matter and dark energy
respectively. An analysis of a combination of SDSS and WMAP data
yields \cite{SDSS} $\Omega_m\approx0.30\pm0.04(1\sigma)$. Crudely
speaking and easy to remember
$$\Omega_b\approx 0.05, \Omega _{CDM}\approx 0.30, \Omega_{\Lambda}\approx 0.65$$
%The combined MAXIMA-1 \cite{MAXIMA-1}, BOOMERANG \cite{BOOMERANG},
%DASI \cite{DASI} and COBE/DMR Cosmic Microwave Background (CMB)
%observations \cite{COBE} imply that the Universe is flat
%\cite{flat01}, $\Omega=1.11\pm0.07$ and that most of the matter in the Universe is Dark, i.e.
%exotic. This has been confirmed by the recent
%WMAP data, which have lead \cite{SPERGEL} to the following results:\\
% $ h=0.72 \pm 0.05$,
%$ \Omega_b~h^2=0.024 \pm 0.001\Rightarrow \Omega_b=0.047\pm 0.06$,
%$ \Omega_m~h^2=0.14 \pm 0.02 \Rightarrow \Omega_m=0.29\pm 0.07$.
%Combining the WMAP data with other experiments one finds: $
%h=0.71^{+0.04}_{-0.03}$,
%  $ \Omega_b~h^2=0.0224 \pm 0.0009\Rightarrow  \Omega_b=0.044\pm 0.04$,
%$ \Omega_m~h^2=0.135^{+0.008}_{-0.009}\Rightarrow \Omega_m=0.27\pm
%0.04$, $ \Omega_{\nu}~h^2<0.0076~(95\%~C.L)\Rightarrow \Sigma
%m_{\nu i}<0.69~eV$, $ \Omega_{\Lambda}=0.69\pm0.08$. In a more
%recent analysis of a combination of SDSS and WMAP data yields
%\cite{SDSS} $\Omega_m\approx0.30\pm0.04(1\sigma)$.  Crudely
%speaking and in a way easy to remember:
%$$\Omega_b=0.05, \Omega _{CDM}= 0.30, \Omega_{\Lambda}= 0.65$$
% The "geometry of the Universe is described \cite{SPERGEL} by:
%$$\Omega=1.02\pm0.02~,~w<-0.78~(95\%~(CL)~(w=-1~for~\Lambda),$$
% while the age of the Universe is \cite{SPERGEL}: $13.7\pm 0.2~Gyr$

Since the non exotic component cannot exceed $40\%$ of the CDM
~\cite {Benne}, there is room for exotic WIMP's (Weakly
Interacting Massive Particles).
  In fact the DAMA experiment ~\cite {BERNA2} has claimed the observation of one signal in direct
detection of a WIMP, which with better statistics has subsequently
been interpreted as a modulation signal \cite{BERNA1}.These  data,
however, if they are due to the coherent process, are not
consistent with other recent experiments, see e.g. EDELWEISS and
CDMS \cite{EDELWEISS}. It could still be interpreted as due to the
spin cross section, but with a new interpretation of the extracted
nucleon cross section.

 Supersymmetry naturally provides candidates for the dark matter constituents
\cite{Jung},\cite{GOODWIT}-\cite{ELLROSZ}.
 In the most favored scenario of supersymmetry the
LSP can be simply described as a Majorana fermion, a linear
combination of the neutral components of the gauginos and
higgsinos \cite{Jung},\cite{GOODWIT}-\cite{ref2}. In most
calculations the neutralino is assumed to be primarily a gaugino,
usually a bino. Models which predict a substantial fraction of
higgsino lead to a relatively large spin induced cross section due
to the Z-exchange. Such models have been less popular, since they
tend to violate the relic abundance constraint.
% The upper bound on this constraint has, however, been decreased by the recent
%WMAP data. In fact  the LSP relic abundance (including
%co-annihilation) is:
%\begin{itemize}
%\item Before WMAP:
%  $$0.09 \leq\Omega_{LSP}h^2\leq0.22$$
%\item After WMAP:
%$$0.09 \leq\Omega_{LSP}h^2\leq0.124$$
%\end{itemize}
These  fairly stringent constrains, however, apply only in the
thermal production mechanism. Furthermore they do not affect the
LSP density in our vicinity derived from the rotational curves.
 We thus feel free to explore the consequences of two recent models \cite{CHATTO},
\cite{WELLS}, which are non-universal gaugino mass models and give
rise to large higgsino components. Sizable spin cross sections
also arise in the context of other models, which have appeared
recently \cite{JDVSPIN04}, \cite{EOSS04}-\cite{HMNS05} (see also
Ellis {\it et al}
\cite{EOSS05} for a recent update of such calculations).\\
 Knowledge of the spin induced nucleon cross section is very
 important since it may lead to transitions to excited states,
 which provide the attractive signature of detecting the
 de-excitation $\gamma$ rays in or without coincidence with the
 recoiling nucleus \cite{eji93}-\cite{VQS04}. Furthermore it may dominate in very light
 systems like the $^3$He, which offer attractive advantages \cite{SANTOS04}.

\section{The Essential Theoretical Ingredients  of Direct Detection.}
 Even though there exists firm indirect evidence for a halo of dark matter
 in galaxies from the
 observed rotational curves, it is essential to directly
detect \cite{Jung},\cite{GOODWIT}-\cite{KVprd}
 such matter. Such a direct detection, among other things, may also
 unravel the nature of the constituents of dark matter. The
 possibility of such detection, however, depends on the nature of its
 constituents. Here we will assume that such a constituent is the
 lightest supersymmetric particle or LSP.
  Since this particle is expected to be very massive, $m_{\chi} \geq 30 GeV$, and
extremely non relativistic with average kinetic energy $T \approx
50KeV (m_{\chi}/ 100 GeV)$, it can be directly detected
~\cite{Jung}-\cite{KVprd} mainly via the recoiling of a nucleus
(A,Z) in elastic scattering. The event rate for such a process can
be computed from the following ingredients:
\begin{enumerate}
\item An effective Lagrangian at the elementary particle (quark)
level obtained in the framework of supersymmetry as described ,
e.g., in Refs~\cite{ref2,JDV96}.
\item A well defined procedure
for transforming the amplitude obtained using the previous
effective Lagrangian from the quark to the nucleon level, i.e.
\beq
{\it L}_{eff} = - \frac {G_F}{\sqrt 2} ({\bar \chi}_1 \gamma^{\lambda}
\gamma^5 \chi_1) J_{\lambda}(Z)
 \label{eq:eg 15}
\eeq
where
\beq
J_{\lambda}(Z) = {\bar N} \gamma_{\lambda} [f^0_V(Z) + f^1_V(Z) \tau_3
+  f^0_A(Z) \gamma_5 + f^1_A(Z) \gamma_5  \tau_3] N
\label{eq:eg 16}
\eeq
The superscripts $0(1)$ refer to the isoscalar (isovector) components of the current. This form
depends, of course, on the quark model for the nucleon. This step is not trivial, since the
obtained results depend crucially on the content of the nucleon in
quarks other than u and d. This is particularly true for the
scalar couplings, which are proportional to the quark
masses~\cite{Dree}$-$\cite{Chen} as well as the isoscalar axial
coupling. \item Nuclear matrix elements.
\cite{Ress}$-$\cite{DIVA00} obtained with as reliable as possible
many body nuclear wave functions. Fortunately in the most studied
case of the scalar coupling the situation is quite simple, since
then one needs only the nuclear form factor. Some progress has
also been made in obtaining reliable static spin matrix elements
and spin response functions \cite{Ress}$-$\cite{DIVA00}.
\end{enumerate}
Since the obtained rates are very low, one would like to be able
to exploit the modulation of the event rates due to the earth's
revolution around the sun \cite{DFS86,FFG88}
\cite{Verg98}$-$\cite{Verg01}. In order to accomplish this one
adopts a folding procedure, i.e one has to assume some velocity
distribution
\cite{DFS86,COLLAR92},~\cite{Verg99,Verg01},\cite{UK01}-\cite{GREEN02}
for the LSP. In addition one would like to exploit the signatures
expected to show up in directional experiments, by observing the
nucleus in a certain direction. Since the sun is moving with
relatively high velocity with respect to the center of the galaxy,
one expects strong correlation of such observations with the
motion of the sun \cite {ref1,UKDMC}. On top of this one expects
to see a more interesting pattern of modulation as well.

 The calculation of this cross section  has become pretty standard.
 One starts with
representative input in the restricted SUSY parameter space as
described in the literature for the scalar interaction~\cite{Gomez,ref2}
 (see also Arnowitt
and Dutta \cite{ARNDU}).
 We will only outline here some features entering the
spin contribution. The spin contribution comes mainly via the
Z-exchange diagram, in which case the amplitude is proportional to
$Z^2_3-Z^2_4$ ($Z_3,Z_4$ are the Higgsino components in the
neutralino). Thus in order to get a substantial contribution the
two higgsino components should be large and different from each
other. Normally the allowed parameter space is constrained so that
the neutralino (LSP) is  primarily gaugino, to allow neutralino
relic abundance in the allowed WMAP region mentioned above. Thus
one cannot take advantage of the small Z mass to obtain large
rates. Models with higgsino-like LSP are possible, but then, as we
have mentioned, the LSP annihilation cross section gets enhanced
and the relic abundance $\Omega_{\chi}~h^2$ gets below the allowed
limit.
 It has recently been shown, however, that in the hyperbolic branch
 of the allowed parameter space \cite{CCN03}, \cite{CHATTO} even
 with a higgsino like neutralino the WMAP relic abundance constraint can be
 respected. So, even though the issue may not be satisfactorily settled,
  we feel that it is worth exploiting the spin cross section in the
 direct neutralino detection, since, among other things, it may populate
 excited  nuclear states, if they happen to be so low in energy that they become
 accessible to  the low energy neutralinos  \cite{eji93}-\cite{VQS04}.

 \section{Rates}
The differential non directional  rate can be written as
\begin{equation}
dR_{undir} = \frac{\rho (0)}{m_{\chi}} \frac{m}{A m_N}
 d\sigma (u,\upsilon) | {\boldmath \upsilon}|
\label{2.18}
\end{equation}
where $d\sigma(u,\upsilon )$ was given above,
$\rho (0) = 0.3 GeV/cm^3$ is the LSP density in our vicinity,
 m is the detector mass and
 $m_{\chi}$ is the LSP mass

 The directional differential rate, in the direction $\hat{e}$ of the
 recoiling nucleus, is \cite{JDVSPIN04}:
\beq
dR_{dir} = \frac{\rho (0)}{m_{\chi}} \frac{m}{A m_N}
|\upsilon| \hat{\upsilon}.\hat{e} ~\Theta(\hat{\upsilon}.\hat{e})
 ~\frac{1}{2 \pi}~
d\sigma (u,\upsilon\
\nonumber \delta(\frac{\sqrt{u}}{\mu_r \upsilon
\sqrt{2}}-\hat{\upsilon}.\hat{e})
 \label{2.20}
\eeq

%\begin{eqnarray}
%dR_{dir} &=& \frac{\rho (0)}{m_{\chi}} \frac{m}{A m_N}
%|\upsilon| \hat{\upsilon}.\hat{e} ~\Theta(\hat{\upsilon}.\hat{e})
% ~\frac{1}{2 \pi}~
%d\sigma (u,\upsilon)\\
%\nonumber & &\delta(\frac{\sqrt{u}}{\mu_r \upsilon
%b\sqrt{2}}-\hat{\upsilon}.\hat{e}) ~~,~ \Theta (x)= \left \{
%\begin{array}{c}1~,x>0\\0~,x<0 \end{array} \right \}
% \label{2.20}
%
%\end{eqnarray}
where $\Theta(x)$ is the Heaviside function and:
\beq
d\sigma (u,\upsilon)== \frac{du}{2 (\mu _r b\upsilon )^2}
 [(\bar{\Sigma} _{S}F(u)^2
                       +\bar{\Sigma} _{spin} F_{11}(u)]
\label{2.9}
\end{equation}
where $ u$ the energy transfer $Q$ in dimensionless units given by
\begin{equation}
 u=\frac{Q}{Q_0}~~,~~Q_{0}=[m_pAb]^{-2}=40A^{-4/3}~MeV
\label{defineu}
\end{equation}
 with  $b$ is the nuclear (harmonic oscillator) size parameter. $F(u)$ is the
nuclear form factor and $F_{11}(u)$ is the spin response function associated with
the isovector channel.
% It is the area of constructing such form factors and momentum
%distributions that Professor M. Grypeos, honored in this volume,
% has made a great contribution \cite{grypeos}.

The scalar
cross section is given by:
\begin{equation}
\bar{\Sigma} _S  =  (\frac{\mu_r}{\mu_r(p)})^2
                           \sigma^{S}_{p,\chi^0} A^2
 \left [\frac{1+\frac{f^1_S}{f^0_S}\frac{2Z-A}{A}}{1+\frac{f^1_S}{f^0_S}}\right]^2
\approx  \sigma^{S}_{N,\chi^0} (\frac{\mu_r}{\mu_r (p)})^2 A^2
\label{2.10}
\end{equation}
(since the heavy quarks dominate the isovector contribution is
negligible). $\sigma^S_{N,\chi^0}$ is the LSP-nucleon scalar cross section.
The spin Cross section is given by:
\begin{equation}
\bar{\Sigma} _{spin}  =  (\frac{\mu_r}{\mu_r(p)})^2
                           \sigma^{spin}_{p,\chi^0}~\zeta_{spin},
\zeta_{spin}= \frac{1}{3(1+\frac{f^0_A}{f^1_A})^2}S(u)
\label{2.10a}
\end{equation}
\begin{equation}
S(u)\approx S(0)=[(\frac{f^0_A}{f^1_A} \Omega_0(0))^2
  +  2\frac{f^0_A}{ f^1_A} \Omega_0(0) \Omega_1(0)+  \Omega_1(0))^2  \, ]
\label{s(u)}
 \end{equation}
 The couplings $f^1_A$ ($f^0_A$) and the nuclear matrix elements $\Omega_1(0)$ ($\Omega_0(0)$) associated
 with the isovector (isoscalar) components are normalized so that, in the case
 of the proton at $u=0$, they yield $\zeta_{spin}=1$.
% The proton and neutron cross sections are related by:
% \beq
% \frac{ \sigma^{spin}_{p,\chi^0}}{ %\sigma^{spin}_{n,\chi^0}}=\frac{(1+\frac{f^0_A}{f^1_A})^2}{(1-\frac{f^0_A}{f^1_A})^2}
 %\label{pncross}
% \eeq
 The proton cross section given by:
 \beq
\sigma^{spin}_{p,\chi^0}=3 \sigma_0 |{f^0_A}+{f^1_A}|^2=3 \sigma_0|a_p|^2~,~
\sigma_0 = \frac{1}{2\pi} (G_F m_p)^2 = 0.77 \times 10^{-38}
cm^2 = 0.77 \times 10^{-2} pb
\label{eq:eg 52}
\eeq
wile for the neutron $\sigma^{spin}_{n,\chi^0}=3 \sigma_0 |{f^0_A}-{f^1_A}|^2=3 \sigma_0|a_n|^2 $.
% In these expressions
% $ f^0_A$, $f^1_A$ are the isoscalar and the isovector axial current couplings at the nucleon level  and
Here $a_p$ and $a_n$ are the usual proton and neutron spin amplitudes \cite{JELLIS}.\\
 With these definitions in the proton neutron representation we get:
 \beq
 \zeta_{spin}= \frac{1}{3}S^{'}(0)
 \label{zeta3}
 \eeq
 \beq
 S^{'}(0)=\left[(\frac{a_n}{a_p}\Omega_n(0))^2+2 \frac{a_n}{a_p}\Omega_n(0) \Omega_p(0)+\Omega^2_p(0)\right]
 \label{Spn}
 \eeq
 where $\Omega_p(0)$ and $\Omega_n(0)$ are the proton and neutron components of the static spin nuclear matrix elements. In extracting limits on the nucleon cross sections from the data we will find it convenient to
 write:
 \begin{equation}
                          \sigma^{spin}_{p,\chi^0}~\zeta_{spin} =\frac{\Omega^2_p(0)}{3}|\sqrt{\sigma_p}+\frac{\Omega_n}{\Omega_p} \sqrt{\sigma_n}
 e^{i \delta}|^2
\label{2.10ab}
\end{equation}
 In Eq. (\ref{2.10ab}) $\delta$ the relative
phase between the two amplitudes defined by
 \beq
 a_N=\sum_{q=u,d,s}d_q \Delta q_N
 \label{amplitudea}
 \eeq
 \beq
 2s_{\mu}\Delta q_N=\left<N|\bar{q}\gamma_{\mu}\gamma_5 q|N\right >
 \label{deltaq}
 \eeq
  where $s_{\mu}$ is the nucleon spin and $d_q$ the relevant spin amplitudes at the quark level obtained in a
  given SUSY model.\\
  The isoscalar and the isovector axial current
couplings at the nucleon level, $ f^0_A$, $f^1_A$, are obtained from the corresponding ones given by the SUSY
 models at the quark level, $ f^0_A(q)$, $f^1_A(q)$, via renormalization
coefficients $g^0_A$, $g_A^1$, i.e.
$ f^0_A=g_A^0 f^0_A(q),f^1_A=g_A^1 f^1_A(q).$
The  renormalization coefficients are given terms of $\Delta q$ defined above \cite{JELLIS},
via the relations
$$g_A^0=\Delta u+\Delta d+\Delta s=0.77-0.49-0.15=0.13~,~g_A^1=\Delta u-\Delta d=1.26$$
We see that, barring very unusual circumstances at the quark level, the isoscalar contribution is
negligible. It is for this reason that one might prefer to work in the isospin basis.
The static spin matrix elements are obtained in the context of a given nuclear model. Some such matrix elements of interest to the planned experiments are given in table \ref{table.spin}.
 The shown results
are obtained from DIVARI \cite{DIVA00}, Ressel {\it et al} (*) \cite{Ress},
 the Finish group (**) \cite {SUHONEN03} and the Ioannina team (+) \cite{ref1}, \cite{KVprd}.

%%%%%%%%%%%%%%%%%%%%%%%%%%%%%%%%%%%%%%%%%%%%%%%%%%%%%%%%%%%%%%%%%%%%%%%%
\begin{table}[t]
\caption{
 The static spin matrix elements for various nuclei. For $^3$He see Moulin, Mayet and Santos
\cite{Santos}. For the other
light nuclei the calculations are from DIVARI \cite{DIVA00}.
 For  $^{73}$Ge and $^{127}$I the results presented  are from Ressel {\it et al}
\cite{Ress} (*) and the Finish group {\it et al} \cite {SUHONEN03}
 (**).
 For $^{207}$Pb they were obtained by the Ioannina team (+).
\cite{ref1}, \cite{KVprd}.
 \label{table.spin} }
\begin{center}
\begin{tabular}{lrrrrrrrr}
\hline\hline
 &   &  &  &  &   &  & &\\
 &$^3$ He& $^{19}$F & $^{29}$Si & $^{23}$Na  & $^{73}$Ge & $^{127}$I$^*$ & $ ^{127}$I$^{**}$ & $^{207}$Pb$^+$\\
\hline
    &   &  &  &  &    \\
$\Omega_{0}(0)$ &1.244     & 1.616   & 0.455  & 0.691  &1.075 & 1.815  &1.220  & 0.552\\
$\Omega_{1}(0)$&-1.527     & 1.675  & -0.461  & 0.588 &-1.003 & 1.105  &1.230  & -0.480\\
$\Omega_{p}(0)$ &-0.141    & 1.646  & -0.003  & 0.640  &0.036 &1.460   &1.225  & 0.036\\
$\Omega_{n}(0)$ &1.386     & -0.030   & 0.459  & 0.051  &1.040 & 0.355  &-0.005 & 0.516\\
$\mu_{th} $& &2.91   &-0.50  & 2.22  &    & &\\
$\mu_{exp}$& &2.62   &-0.56  & 2.22  &    & &\\
$\frac{\mu_{th}(spin)}{ \mu_{exp}}$& &0.91   &0.99  & 0.57  &    &  &\\
\hline
\hline
\end{tabular}
\end{center}
\end{table}
%%%%%%%%%%%%%%%%%%%%%%%%%%%%%%%%%%%%%%%%%%%%%%%%%%%%%%%%%
The spin ME are defined as follows:
\beq
\Omega_p(0)=\sqrt{\frac{J+1}{J}}\prec J~J| \sigma_z(p)|J~J\succ ~~,~~
\Omega_n(0)=\sqrt{\frac{J+1}{J}}\prec J~J| \sigma_z(n)|J~J\succ
\label{Omegapn}
\eeq
where $J$ is the total angular momentum of the nucleus and $\sigma_z=2 S_z$. The spin operator is defined by $S_z(p)=\sum_{i=1}^{Z} S_z(i)$, i.e. a sum over all protons in the nucleus,  and
$S_z(n)=\sum_{i=1}^{N}S_z(i)$, i.e. a sum over all neutrons. Furthermore
\beq
\Omega_0(0)=\Omega_p(0)+\Omega_n(0)~~,~~
\Omega_1(0)=\Omega_p(0)-\Omega_n(0)
\label{Omegaiso}
\eeq
\section{Expressions for the Rates}
To obtain the total rates one must fold with LSP velocity and integrate  the
above expressions  over the
energy transfer from $Q_{min}$ determined by the detector energy cutoff to $Q_{max}$
determined by the maximum LSP velocity (escape velocity, put in by hand in the
Maxwellian distribution), i.e. $\upsilon_{esc}=2.84~\upsilon_0$ with  $\upsilon_0$
the velocity of the sun around the center of the galaxy($229~Km/s$).

%\subsection{Non directional rates}
Ignoring the motion of the Earth the total (non directional) rate
is given by
\begin{equation}
R =  \bar{R}\, t(a,Q_{min}) \, \label{3.55f}
\end{equation}
$$ \bar{R}=\frac{\rho (0)}{m_{\chi^0}} \frac{m}{Am_p}~
              (\frac{\mu_r}{\mu_r(p)})^2~ \sqrt{\langle
v^2 \rangle } [\sigma_{p,\chi^0}^{S}~A^2+
 \sigma _{p,\chi^0}^{spin}~\zeta_{spin}]$$
 The SUSY parameters have been absorbed in $\bar{R}$. The
 parameter $t$ takes care of the nuclear form factor and the
 folding with LSP velocity distribution \cite{Verg00,Verg01,JDVSPIN04}
 (see table \ref{table.murt}). It depends on
$Q_{min}$, i.e.  the  energy transfer cutoff imposed by the
detector and $a=[\mu_r b \upsilon _0 \sqrt 2 ]^{-1}$.\\
  In the present work  we find it convenient to re-write it as:
\begin{equation}
R= \bar{K} \left[c_{coh}(A,\mu_r(A))\frac{ \sigma_{p,\chi^0}^{S}}{\sigma_1}+
c_{spin}(A,\mu_r(A))\frac{\sigma _{p,\chi^0}^{spin}}{\sigma_1}~\zeta_{spin} \right]
\label{snew}
\end{equation}
 with $\sigma _1=10^{-5}pb$ and
\beq
\bar{K}=\frac{\rho (0)}{100\mbox{ GeV}} \frac{m}{m_p}~
              \sqrt{\langle v^2 \rangle }~\sigma_1 \simeq 1.60~10^{-2}~ y^{-1}\frac{\rho(0)}{0.3GeVcm^{-3}}
\frac{m}{1Kg}\frac{ \sqrt{\langle
v^2 \rangle }}{280kms^{-1}}
\label{Kconst}
\eeq
and
\begin{equation}
c_{coh}(A, \mu_r(A))=\frac{100\mbox{ GeV}}{m_{\chi^0}}\left[ \frac{\mu_r(A)}{\mu_r(p)} \right]^2 A~t_{coh}(A)~,
c_{spin}(A, \mu_r(A))=\frac{100GeV}{m_{\chi^0}}\left[ \frac{\mu_r(A)}{\mu_r(p)} \right]^2 \frac{t_{spin}(A)}{A}
\label{ctm}
\end{equation}
The parameters $c_{coh}(A,\mu_r(A))$, $c_{spin}(A,\mu_r(A))$, which give the relative merit
 for the coherent and the spin contributions in the case of a nuclear
target compared to those of the proton,  are tabulated in table
\ref{table.murt}
 for
energy cutoff $Q_{min}=0,~10$ keV.\\
Via  Eq. (\ref{snew}) we can  extract the nucleon cross section from
 the data.\\
We distinguish the following cases:
\begin{itemize}

\item If the nuclear contribution
 comes predominantly from protons
 ($\Omega_1=\Omega_0=\Omega_p$), $S(u)\approx\Omega_p^2$ and
\beq \zeta_{spin}=\frac{\Omega^2_p}{3} \label{zeta1} \eeq So
knowing the nuclear matrix element one can extract from the data
the proton cross section. One such example with negligible neutron
contribution is $^{19}$F (see table \ref{table.spin})
%The neutron cross section is related to the proton cross section via Eq.
%\ref{pncross}. So one needs the ratio of the isoscalar to isovector elementary amplitudes.
\item If the nuclear contribution comes predominantly
from neutrons (($\Omega_0=-\Omega_1=\Omega_n$), one can similarly extract the neutron cross
section from the data.
%One can also extract the proton cross section, if the ratio of the
%isoscalar to isovector amplitude is known since:
%$$ \zeta_{spin}=\frac{\Omega^2_n}{3} \frac{(1-\frac{f^0_A}{f^1_A})^2}{(1+\frac{f^0_A}{f^1_A})^2}$$
\item In many cases, however, the nuclear structure is such that
one can have contributions from both protons and neutrons. The
situation is then complicated and will be discussed below (see  exclusion plots below).
%sec. \ref{exclplots}).
\\We have seen, however, that in going from the quark to the nucleon level we encounter the renormalization factors $g^0_A=0.1~~,~~g^1_A=1.2$. Thus the
isoscalar contribution
 is suppressed and
 $$S(0)\approx \Omega^2_1~,~\zeta_{spin}=\frac{\Omega^2_1}{3}$$
Then the proton and the neutron  spin cross sections are the same.
\end{itemize}

\begin{figure}
\begin{center}
\includegraphics[height=.20\textheight]{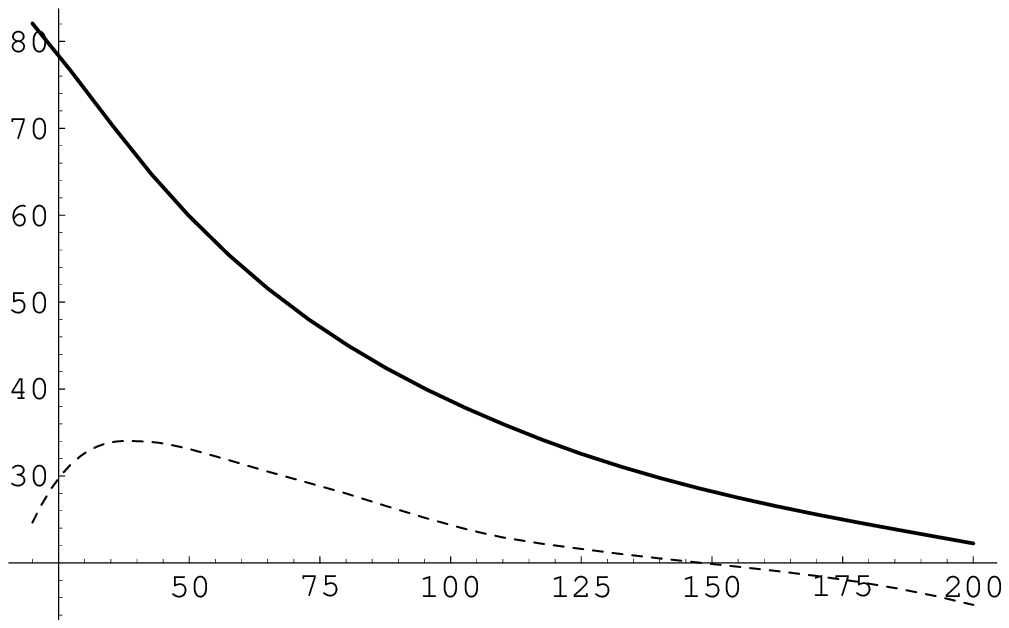}
\includegraphics[height=.20\textheight]{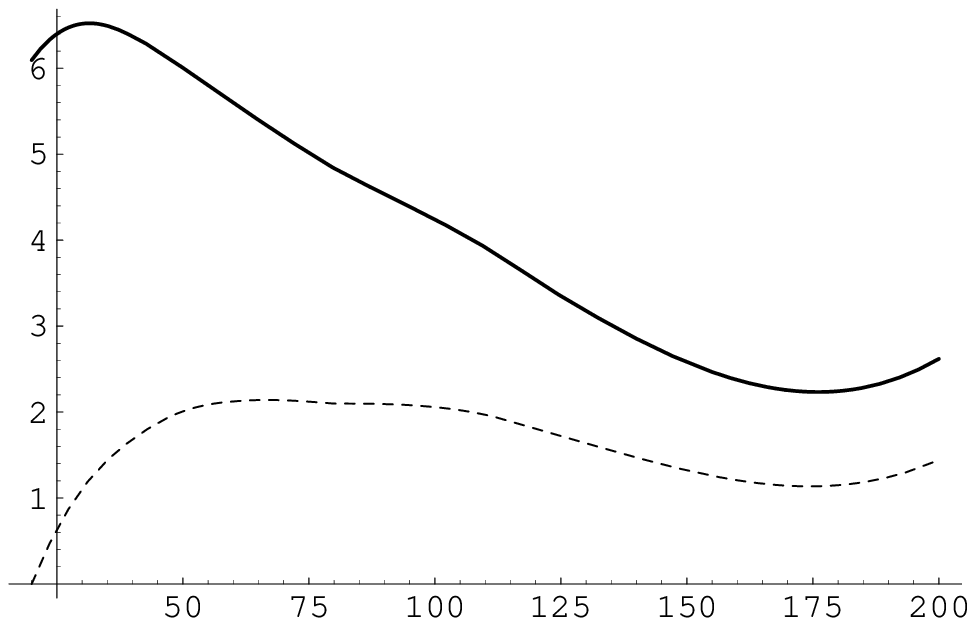}
\caption{ On the left the quantity $c_{spin}(A,\mu_r(A))~\zeta_{spin}$ for the A=19 system is
 shown for two cut off values $Q_{min}=0$,
continuous curve,
and $Q_{min}=10$ keV, dotted curve. On the right the same quantity is shown for the A=127 system.
The advantages of the lighter target, especially for light LSP, are obvious.
 \label{nume} }
\end{center}
\end{figure}

 Neglecting the isoscalar contribution and using
$\Omega^2_1=1.22$ and $\Omega^2_1=2.8$ for $^{127}$I and $^{19}$F respectively
the extracted nucleon cross sections satisfy:
\begin{equation}
\frac{\sigma^{spin}_{p,\chi^0}}{\sigma^{S}_{p,\chi^0}} =
 \left[\frac{c_{coh}(A,\mu_r(A))}{c_{spin}(A,\mu_r(A))}\right]
\frac{3 }{\Omega^2_1} \Rightarrow
\approx  \times 10^{4}~(A=127)~,~ \approx \times 10^2~(A=19)
\label{ratior2}
\end{equation}
% While for $^{19}$F ($\Omega^2_1=2.8$) we get:
%\begin{equation}
%\frac{\sigma^{spin}_{p,\chi^0}}{\sigma^{S}_{p,\chi^0}} \approx 1\times 10^2
%\label{ratio2b}
%\end{equation}
It is for this reason that the limit on the spin proton cross section extracted from both
targets is much poorer.
% For heavy LSP, $\ge 100$ GeV, due to the nuclear form factor, $t_{spin}(127)< t_{spin}(19)$.  
% This disadvantage cannot be overcome by the larger reduced mass. It even becomes worse,
%  if the effect of the spin ME is included. 
%%%%%%%%%%%%%%%%%%%%%%%%%%%%%%%%%%%%%%%%%%%%%%%%%%%%%%%%%%%%%%%%%%%%%%%%%%
\begin{table}[t]
\caption{
The factors $c19= c_{coh}(19,\mu_r(19))$,  $s19= c_{spin}(19,\mu_r(19))$,
$c19= c_{coh}(73,\mu_r(73))$,  $s73= c_{spin}(73,\mu_r(73))$
 and
$c127=c_{coh}(127,\mu_r(127))$,  $s127= c_{spin}(127,\mu_r(127))$
for two values of $Q_{min}$. Also given are the factors $s3= c_{spin}(3,\mu_r(3))$ for  $Q_{min}=0$.
\label{table.murt}
}
\begin{center}
%{\footnotesize
\begin{tabular}{|r|r|rrrrrrrr|}
%\hline
\hline
$Q_{min}$& &\multicolumn{8}{c|}{$m_{\chi}$ (GeV)}\\
\hline
& &   &  &  &  &   & & &\\
keV& & 20 & 30 & 40  & 50 & 60 & 80&100&200\\
\hline
0&t(3,s)&2.3&2.3&2.3&2.3&2.3&2.3&2.3&2.3\\
\hline
0&s3&29.1& 20.6& 15.9& 12.9& 10.9& 8.3& 6.7&  3.4\\
\hline
\hline
0&t(19,c)&1.153&1.145&1.138&1.134&1.130&1.124&1.121&1.112\\
0&t(19,s)&1.132&1.117&1.105&1.096&1.089&1.079&1.072&1.056\\
\hline
0&c19&11465& 10478& 9423& 8499& 7702& 6451& 5539& 3212\\
0&s19&31.2& 28.3& 25.4& 22.8& 20.6& 17.2& 14.6& 8.4\\
\hline
0&t(73,c)& 2.238& 2.166& 2.094& 2.028& 1.967& 1.865& 1.785& 1.559\\
0&t(73,s)& 2.270& 2.223& 2.175& 2.129& 2.086& 2.012& 1.952&1.771\\
\hline
0&c73&225512& 261081& 278461& 284755& 284569& 274313&258838 & 186743\\
0&s73&41.6& 47.7& 50.3& 50.9& 50.4& 47.7& 44.4& 30.8\\
\hline
0&t(127,c)&0.984&0.844&0.721&0.621&0.542&0.430&0.358&0.213\\
0&t(127,s)&0.948&0.796&0.671&0.574&0.501&0.401&0.340&0.220\\
\hline
0&c127& 205674& 224676& 222547& 211216& 196895& 168585& 145173&  82424\\
0&s127&12.3& 13.1& 12.8& 12.1& 11.3& 9.7& 8.5& 5.3\\
\hline
\hline
10&t(19,c)&0.352&0.511&0.592&0.639&0.667&0.710&0.720&0.773\\
10&t(19,s)&0.340&0.489&0.563&0.606&0.631&0.669&0.676&0.720\\
\hline
10&c19&3500& 4676& 4902& 4789& 4546& 4075& 3557& 2233\\
10&s19& 9.3& 12.4& 12.9& 12.6& 11.9& 10.6& 9.3& 5.8\\
\hline
 10&t(73,c)&0& 0.020& 0.119& 0.246& 0.363& 0.539& 0.651& 0.847\\
10&t(73,s)&0& 0.0175& 0.105& 0.213& 0.311& 0.453& 0.539& 0.677\\
\hline
10&c73)&0& 2313& 15295& 32947& 49559& 73463& 86339&  89290\\
10&s73&0& 0.39& 2.5& 5.3& 7.9& 11.6& 13.4& 13.4\\
\hline
10&t(127,c)&0.000&0.156&0.205&0.222&0.216&0.191&0.175&0.109\\
10&t(127,s)&0.000&0.135&0.177&0.192&0.190&0.174&0.165&0.121\\
\hline
10&c127& 0& 41528& 63276& 75507& 78468& 74883& 70964& 42180\\
10&s127& 0.& 2.2& 3.4& 4.0& 4.3& 4.2& 4.1& 2.9\\
\hline
\end{tabular}
%\label{table.murt}
%}
\end{center}
\end{table}
%%%%%%%%%%%%%%%%%%%%%%%%%%%%%%%%%%%%%%%%%%%%%%%%%%%%%%%%%
%\subsection{Modulated Rates.}
\\The  quantity $c_{spin}(A,\mu_r(A))~\zeta_{spin}$, when the isoscalar contribution is
neglected and employing $\Omega ^2_1= 1.22~ (2.81)$ for $^{127}I$
$(^{19}F)$, is shown in Fig \ref{nume}.In the case of the spin induced cross section the light nucleus $^{19}$F has certainly an advantage over
the heavier nucleus $^{127}$I (see Fig. \ref{nume}).
For the coherent process, however, the light nucleus is no match
 (see Table \ref{table.murt}).
\\If the effects of the motion of the Earth around the sun are included, the total
 non directional rate is given by
\begin{equation}
R= \bar{K} \left[c_{coh}(A,\mu_r(A)) \sigma_{p,\chi^0}^{S}(1 +  h(a,Q_{min})cos{\alpha})\right]
\label{3.55j}
\end{equation}
and an analogous one for the spin contribution.  $h$  is the modulation amplitude and
 $\alpha$ is the phase of the Earth, which is
zero around June 2nd. We are not going to elaborate further on this point since amplitude $h$ depends only on
the LSP mass and is independent of the other SUSY parameters.
\section{Some SUSY input}
\begin{figure}
\begin{center}
\includegraphics[height=.20\textheight]{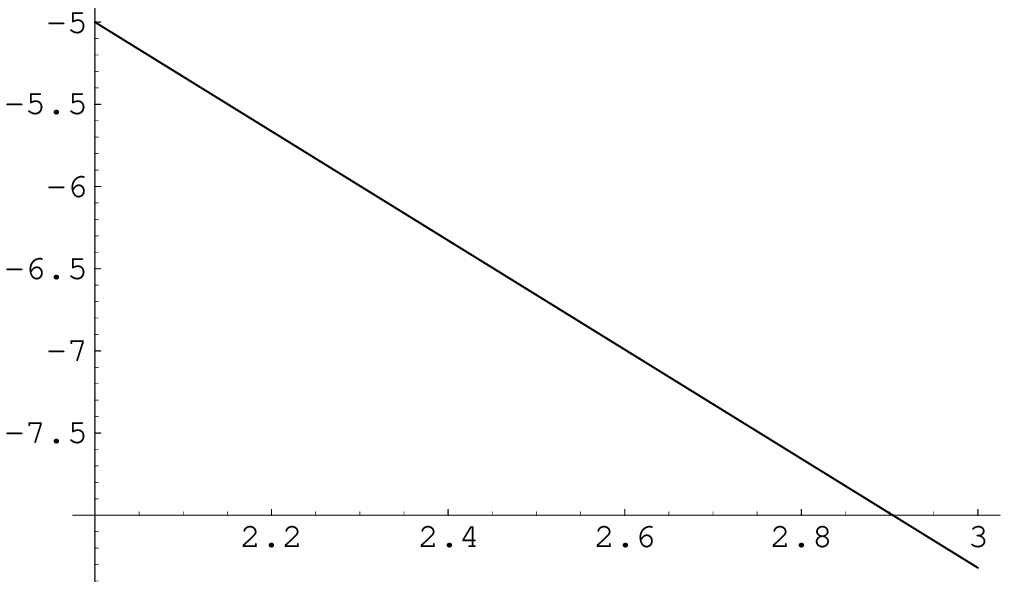}
\includegraphics[height=.20\textheight]{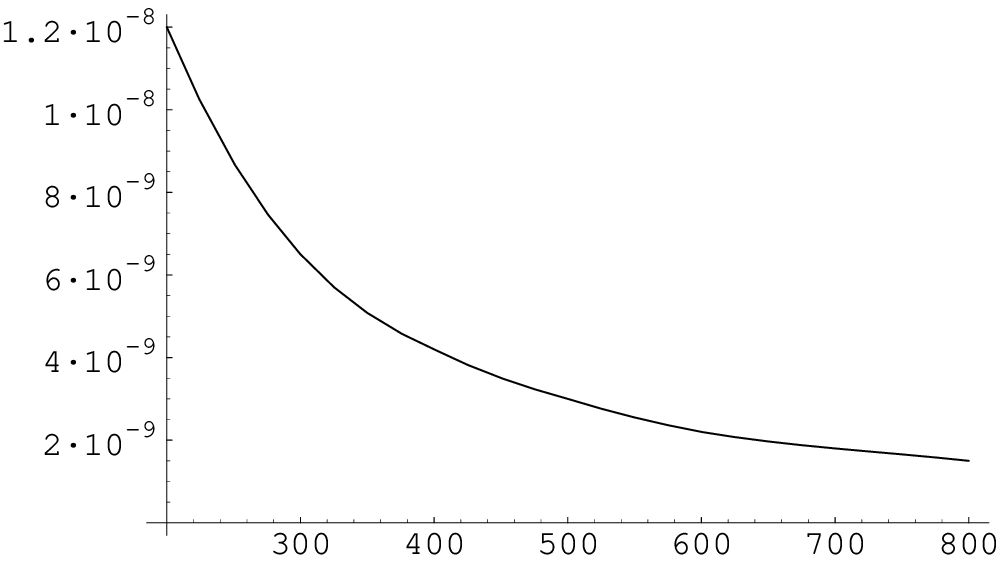}
\caption{ On the left the proton cross section in $pb$ is shown as a
function of the LSP mass (in GeV) obtained by Ellis {\it et al}
\cite{OLIVE04},\cite{EOSS05} (logarithmic scale on both axes). On
the right we show the results obtained by Hisano {\it et al}
\cite{HMNS05}.
 \label{protoncsc} }
\end{center}
\end{figure}

The  most important input towards computing the event
rate is the spin nucleon  cross section. For orientation purposes will only incorporate the results of  two recent calculations:
\begin{itemize}
\item Ellis et al \cite{EOSS04} have
given such cross sections in the allowed SUSY parameter space
taking into account all available experimental constraints and
provided an update of such calculations \cite{EOSS05}. They
have also calculated \cite{OLIVE04} the spin proton cross section as
a function of the LSP mass. Excluding a few data points
we present their results in
Fig. \ref{protoncsc}.\\
\item  In a recent paper Hisano, Matsumoto, Noijiri and Saito \cite{HMNS05}have obtained
 the proton cross section for
Wino and Higgsino like LSP at the one loop level for relatively high LSP mass, where the
 tree contribution is negligible.
Their results for the spin cross section are  shown in the same figure, Fig. \ref{protoncsc}.
\end{itemize}

\begin{figure}
\begin{center}
\includegraphics[height=.20\textheight]{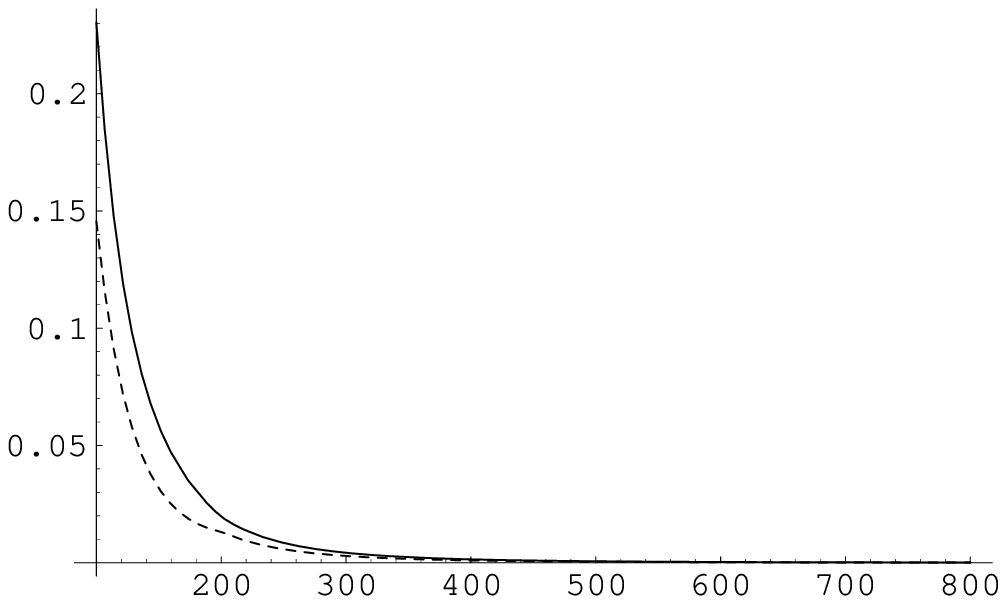}
\includegraphics[height=.20\textheight]{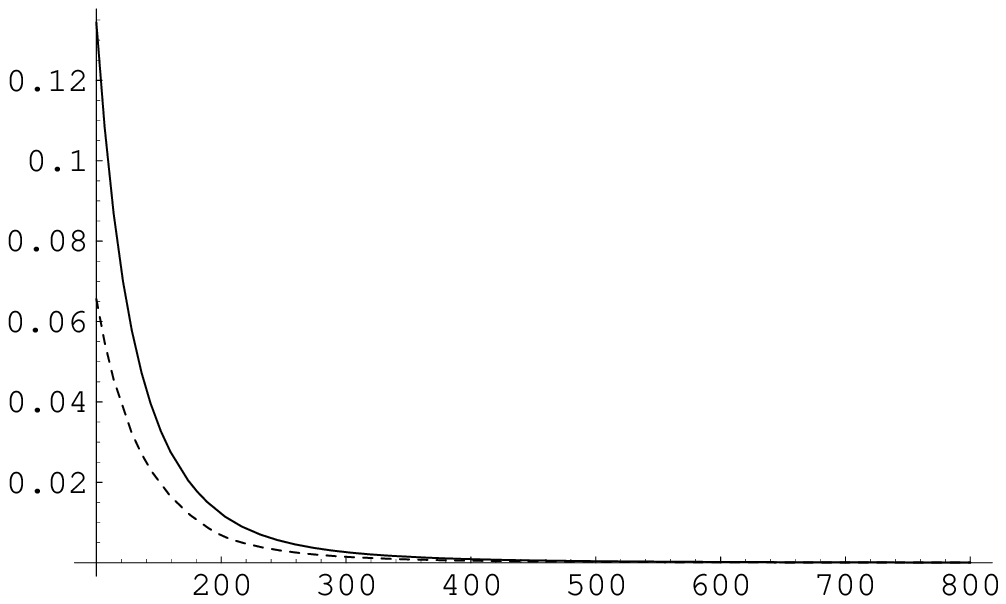}
\caption{ The event  rate per kg-y for the A=19 system on the left
and for the A=127 on the right using the proton cross section of
Ellis {\it et al} \cite{OLIVE04},\cite{EOSS05}. Otherwise the
notation is the same as in Fig. \ref{nume}.
 \label{rateellis} }
\end{center}
\end{figure}

\begin{figure}
\begin{center}
\includegraphics[height=.20\textheight]{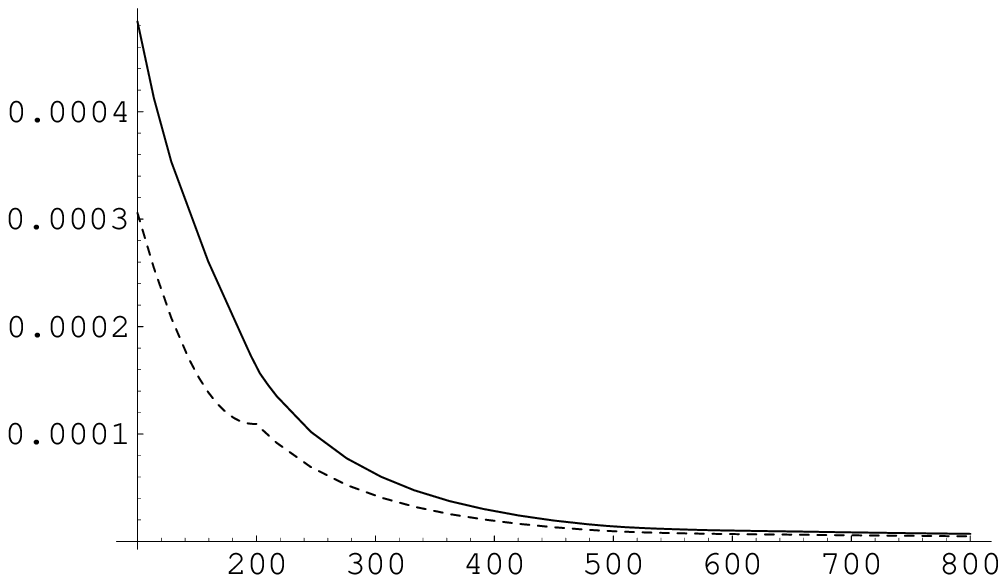}
\includegraphics[height=.20\textheight]{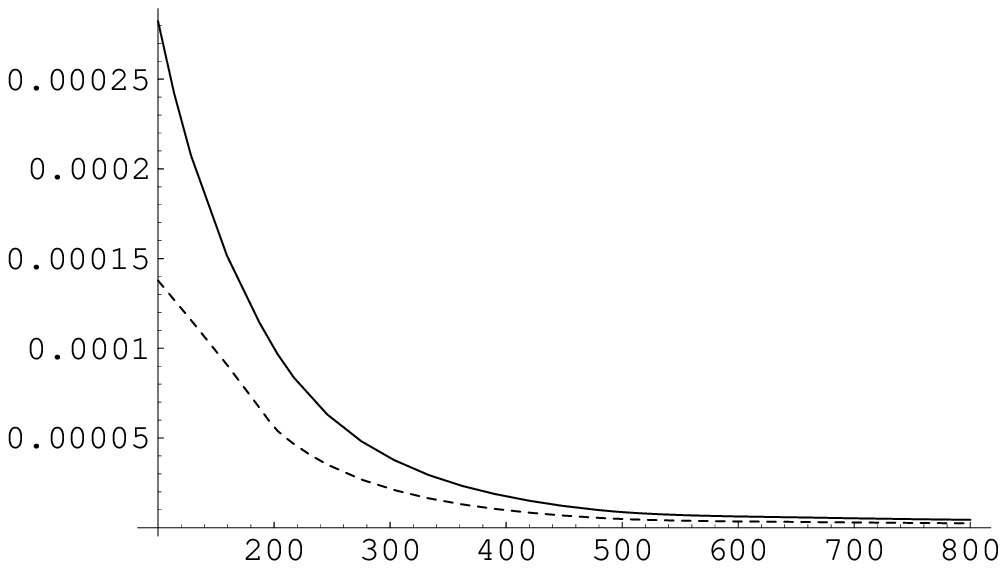}
\caption{ The event rate per kg-y for the A=19 system on the left
and for the A=127 on the right using the proton cross section of
Hisano {\it et al} \cite{HMNS05}. Otherwise the notation is the
same as in Fig. \ref{nume}.
 \label{rate2loop} }
\end{center}
\end{figure}

The event rate obtained with the proton cross section of Ellis
{\it et al} \cite{OLIVE04},\cite{EOSS05} is shown in Fig.
\ref{rateellis}, while for the proton cross section of Hisano {\it et
al} \cite{HMNS05} it is shown in Fig. \ref{rate2loop}.
\section{Bounds on the scalar proton cross section}
\begin{figure}
\begin{center}
\rotatebox{90}{\hspace{0.0cm} $\sigma_p\rightarrow 10^{-5}$pb}
\includegraphics[height=.15\textheight]{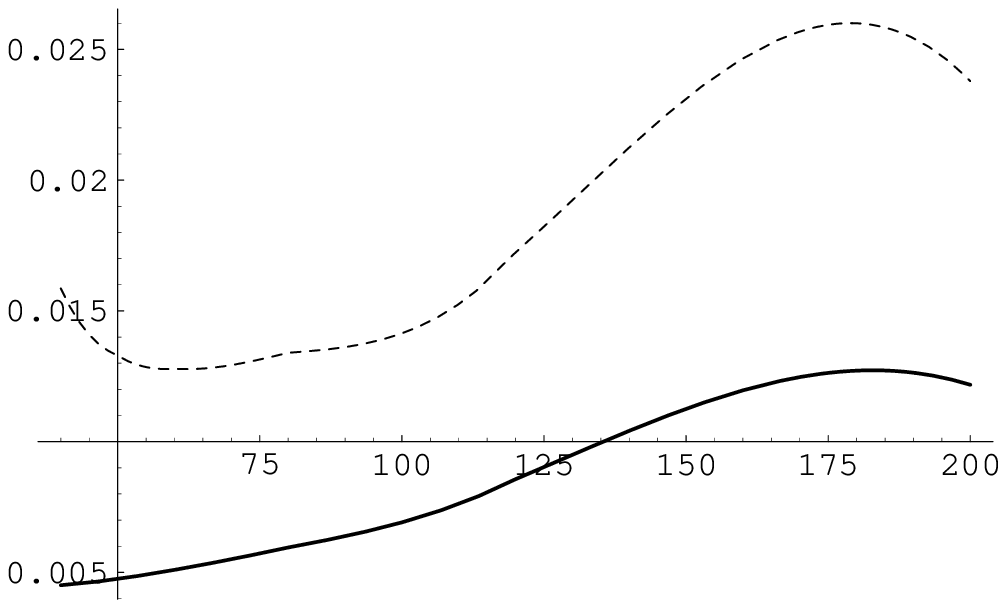}
%\vspace{0.5cm}
%\hspace{-0.5cm}
\hspace{-0.0cm} $m_{\chi}\rightarrow$ GeV
\rotatebox{90}{\hspace{0.0cm} $\sigma_p\rightarrow 10^{-5}$pb}
\includegraphics[height=.15\textheight]{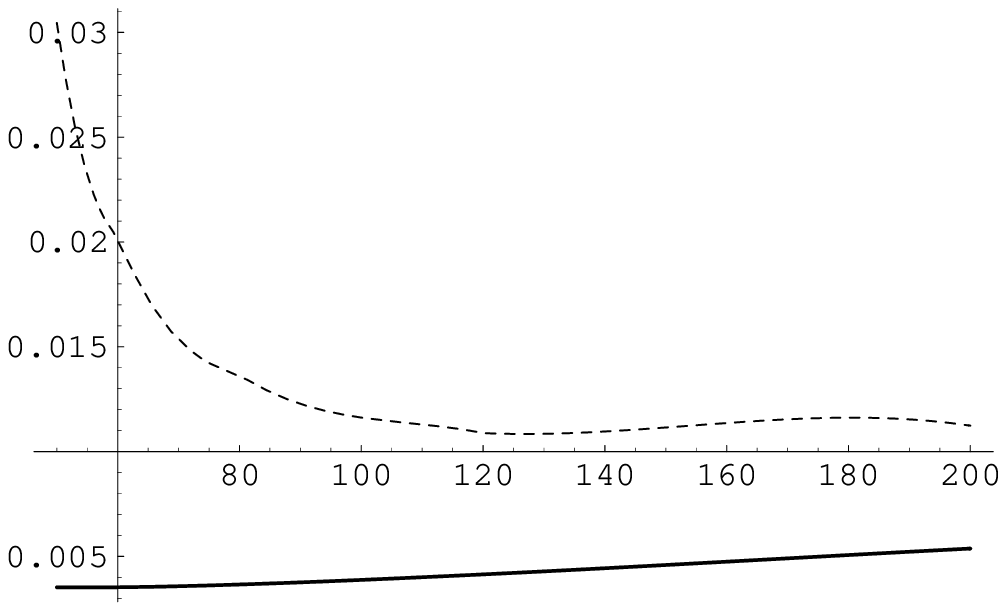}
\hspace{-0.0cm} $m_{\chi}\rightarrow$ GeV
\caption{ The limits on the scalar proton cross section for A$=127$ on the left and A$=73$ on the right as functions of $m_{\chi}$. The continuous (dashed) curves correspond to $Q_{min}=0~(10)$ keV respectively. Note that the advantage of the larger nuclear mass number of the A$=127$ system is counterbalanced by the favorable form factor dependence of the A$=73$ system.
 \label{b127.73}}
\end{center}
\end{figure}

\begin{figure}
\begin{center}
\rotatebox{90}{\hspace{0.0cm} $\sigma_p\rightarrow 10^{-5}$pb}
\includegraphics[height=.15\textheight]{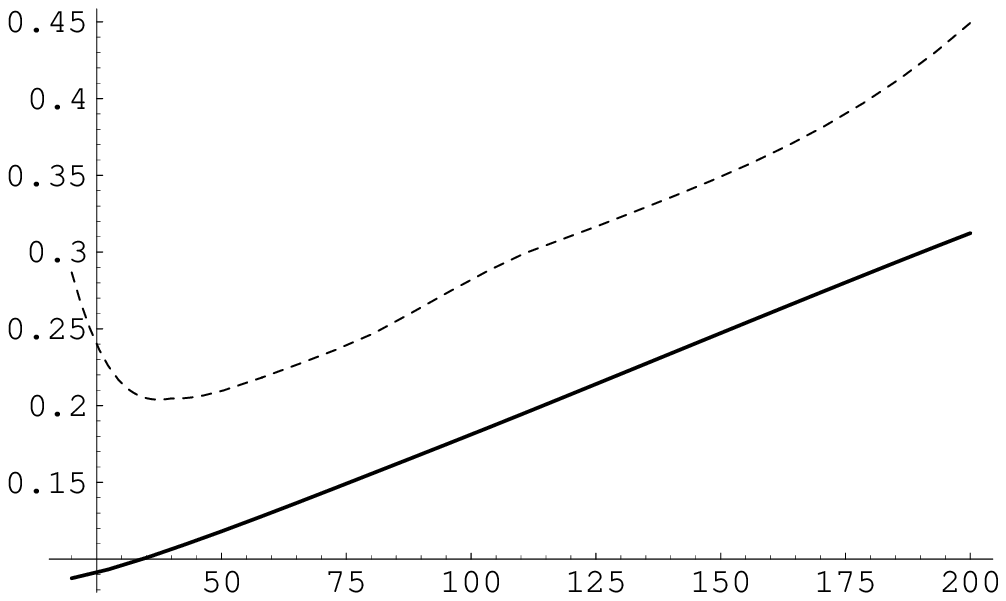}
%\vspace{0.5cm}
%\hspace{-0.5cm}
\hspace{-0.0cm} $m_{\chi}\rightarrow$ GeV
\rotatebox{90}{\hspace{0.0cm} $\sigma_p\rightarrow 10^{-5}$pb}
\includegraphics[height=.15\textheight]{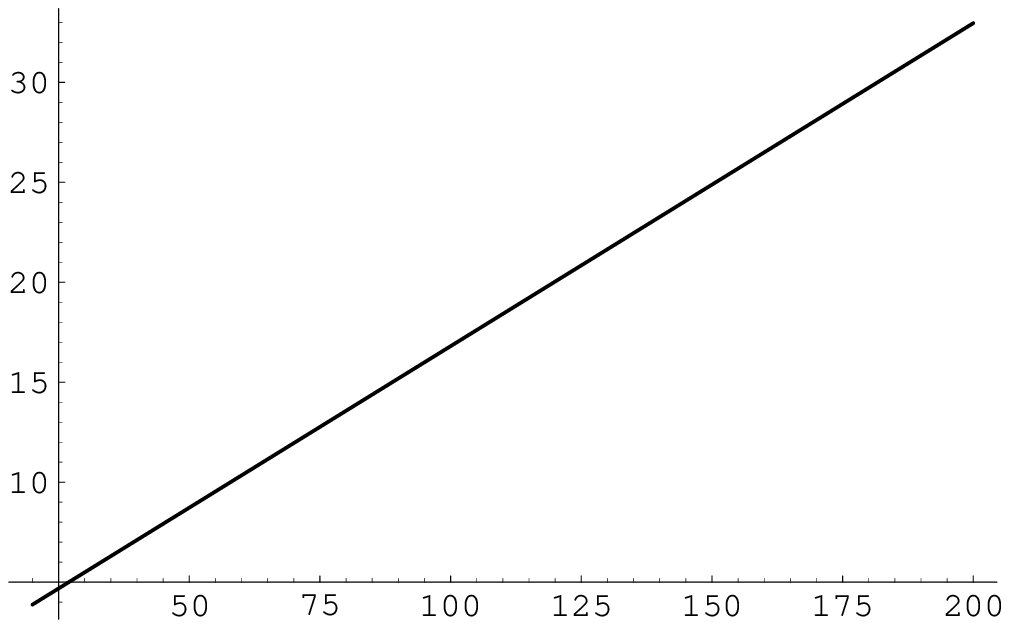}
\hspace{-0.0cm} $m_{\chi}\rightarrow$ GeV
\caption{ The same as in Fig. \ref{b127.73} for A$=19$ on the left and A$=3$ on the right. As expected in the case of the coherent process the light systems are no match against the heavier ones (compare this figure with Fig. \ref{b127.73} )
 \label{b19.3}}
\end{center}
\end{figure}

Before proceeding with the analysis of the spin contribution we would like to discuss the limits on the
scalar proton cross section.
In what follows we will employ for all targets the limit of CDMS II for the Ge target \cite{CDMSII04},
 i.e.  $<2.3$ events for 
an exposure of $52.5$ Kg-d with a threshold of $10$ keV. This event rate is similar to that for other systems \cite{SGF05}. The thus obtained limits are exhibited in Figs \ref{b127.73}-\ref{b19.3}.
\section{Results-Exclusion Plots in the $a_p,a_n$ and $\sigma_p,\sigma_n$ Planes}
\label{exclplots}
\begin{figure}
\begin{center}
\rotatebox{90}{\hspace{0.0cm} $r(m_{\chi},127)\rightarrow$}
\includegraphics[height=.15\textheight]{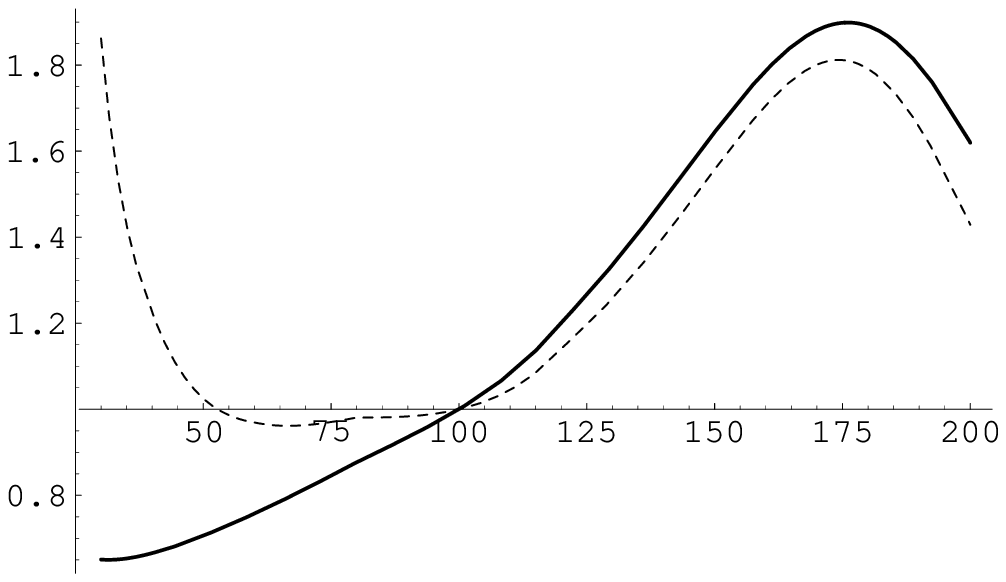}
%\vspace{0.5cm}
%\hspace{-0.5cm}
\hspace{-0.0cm} $m_{\chi}\rightarrow$ GeV
\rotatebox{90}{\hspace{0.0cm} $r(m_{\chi},73)\rightarrow$}
\includegraphics[height=.15\textheight]{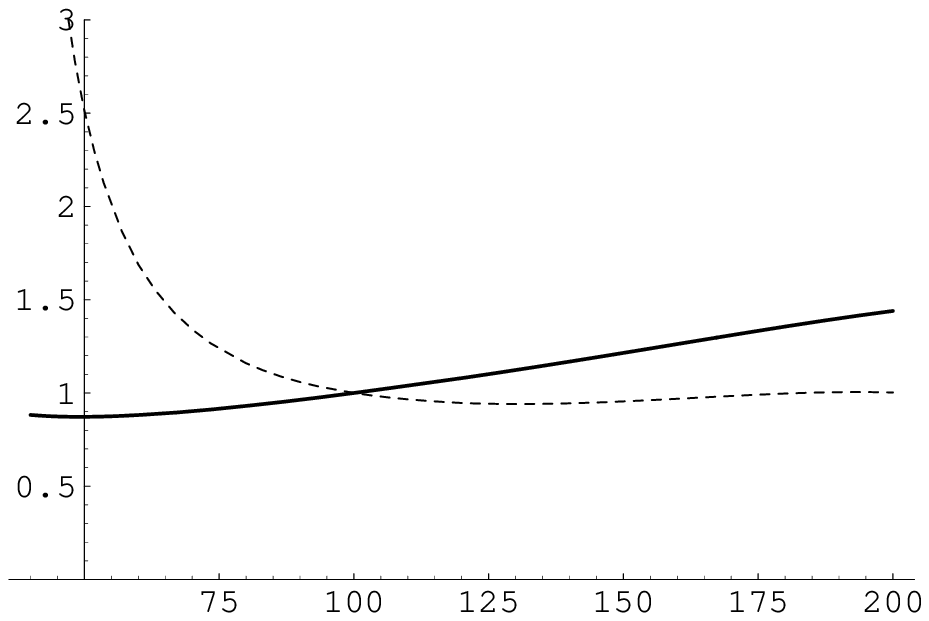}
\hspace{-0.0cm} $m_{\chi}\rightarrow$ GeV
\caption{ The coefficients $r(m_{\chi},127)$ on the left and $r(m_{\chi},73)$ on the right as functions
of $m_{\chi}$. Note that due to the adopted normalization, for sufficiently large $m_{\chi}$ there is little difference between zero threshold (continuous curve) and a threshold of $10$ keV (dashed curve).
 For small LSP mass the results with a threshold of $Q_{min}=10$ keV are not very reliable, since the relevant rate is very small.
 \label{plot.murt127.73}}
\end{center}
\end{figure}

\begin{figure}
\begin{center}
\rotatebox{90}{\hspace{0.0cm} $r(m_{\chi},19)\rightarrow$}
\includegraphics[height=.15\textheight]{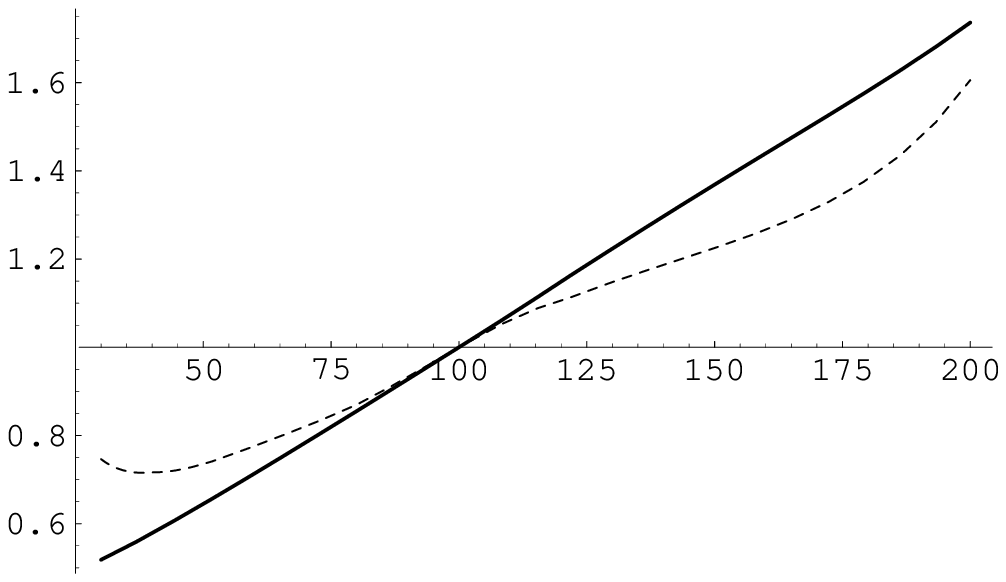}
%\vspace{0.5cm}
%\hspace{-0.5cm}
\hspace{-0.0cm} $m_{\chi}\rightarrow$ GeV
\rotatebox{90}{\hspace{0.0cm} $r(m_{\chi},3)\rightarrow$}
\includegraphics[height=.15\textheight]{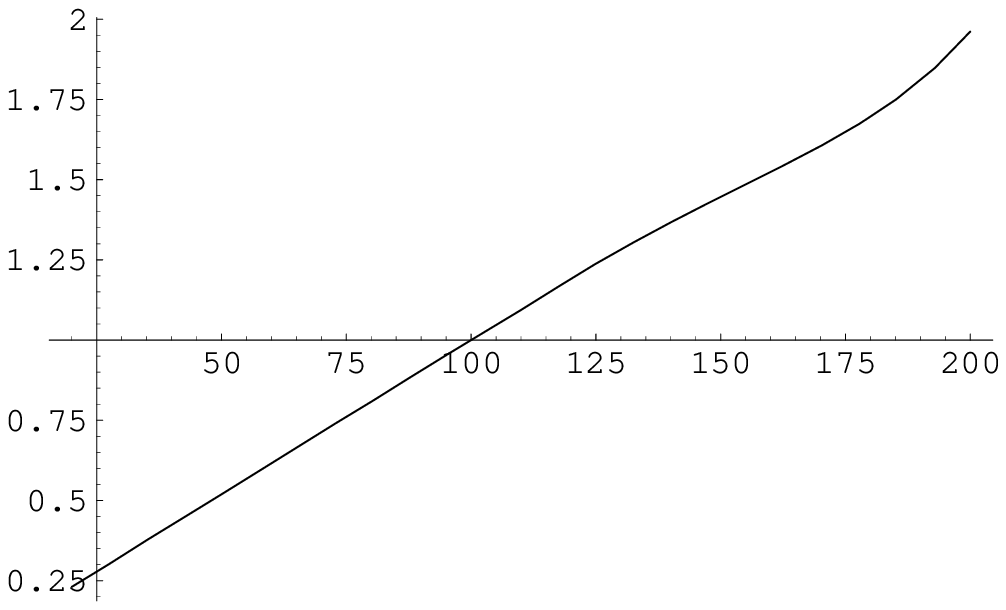}
\hspace{-0.0cm} $m_{\chi}\rightarrow$ GeV
\caption{ The same as in fig \ref{plot.murt127.73} for the coefficients $r(m_{\chi},19)$ on the left and $r(m_{\chi},3)$. Only zero threshold was considered for $^3$He.
\label{plot.murt19.3}}
\end{center}
\end{figure}

 From the data one can extract a
restricted region in the  $\sigma_p,\sigma_n$ plane, which depends
on the event rate and the LSP mass. Some such exclusion plots have already appeared \cite{SGF05}-\cite{GIUGIR05}.
One can plot the constraint imposed on the quantities $|a_p+\frac{\Omega_n}{\Omega_p}a_n|$ and$|\sqrt{\sigma_p}+\frac{\Omega_n}{\Omega_p} \sqrt{\sigma_n}
 e^{i \delta}|^2$ derived from the experimental limits via relations:
 \beq
 |\sqrt{\sigma_p}+\frac{\Omega_n}{\Omega_p} \sqrt{\sigma_n}
 )e^{i \delta}|^2\preceq \sigma _{bound}(A)~r(m_{\chi},A)~;~\sigma _{bound}(A)=\frac{R}{\bar{K}} \frac{3}{\Omega^2_p}\frac{10^{-5} pb}{c^{100}_{spin}(A,\mu_r(A))}~,~r(m_{\chi},A)=\frac{c^{100}_{spin}(A,\mu_r(A))}{c_{spin}(A,\mu_r(A))}
 \label{constraintsigma}
 \eeq
 where $\delta$ is the phase difference between the two amplitudes and ${c^{100}_{spin}(A,\mu_r(A))}$ is the value of ${c_{spin}(A,\mu_r(A))}$ evaluated for the LSP mass of $100$ GeV. Furthermore
 \beq|a_p+\frac{\Omega_n}{\Omega_p}a_n|\preceq  a _{bound}(A)~\left [r(m_{\chi},A) \right ]^{1/2}~,~a _{bound}(A)=\left[\frac{\sigma _{bound}(A)}{3 \sigma_0} \right]^{1/2}
 \label{constraintampl}
 \eeq
 The constraints will be obtained using the functions ${c^{100}_{spin}(A,\mu_r(A))}$, obtained without
 energy cut off , $Q_{min}=0$, even though the experiments have energy cut offs greater than zero.
Furthermore even though we know of no model such that $e^{i\delta}$ is complex, for completeness we will examine below this case as well. Such plots depend on the relative magnitude of the spin matrix elements. They will be given in units of
the A-dependent quantity $\sigma _{bound}(A)$ for the nucleon cross sections and the dimensionless
quantity $a _{bound}$ for the amplitudes respectively. The quantities $r(m_{\chi},A)$, obtained from the data of table \ref{table.murt}, are plotted for $A=127,~73,~19$ and $A=3$ in Figs.
\ref{plot.murt127.73} and \ref{plot.murt19.3}.
 Before we proceed further we should mention  that, if both protons and neutrons contribute, the
standard exclusion plot, must be replaced by a sequence of plots, one for
each LSP mass or via three dimensional plots. We found it is adequate to provide one such plot for a standard LSP mass, e.g. $100$ GeV, and zero energy threshold. The interested reader can deduce the scale for any other case with the help of Figs \ref{plot.murt127.73} and \ref{plot.murt19.3}. In what follows we will employ for all targets the limit  discussed in the previous section.
%of CDMS II for the Ge target \cite{CDMSII04}, i.e.  $<2.3$ events for 
%an exposure of $52.5$ Kg-d with a threshold of $10$ keV. This event rate is similar to that 
%for other systems \cite{SGF05}.

\subsection{ Spin matrix elements of the same sign and $\Omega_p\gg \Omega_n$}
\begin{figure}
\begin{center}
\rotatebox{90}{\hspace{0.0cm} $a_n\rightarrow 0.65$}
\includegraphics[height=.18\textheight]{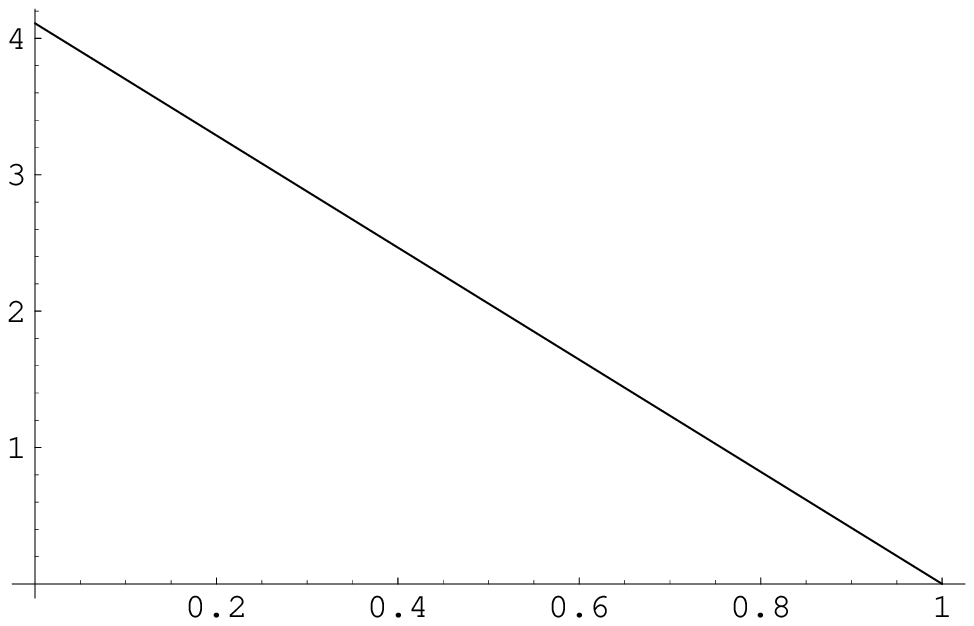}
\hspace{-0.5cm} $a_p\rightarrow 0.65$
\rotatebox{90}{\hspace{0.0cm} $a_n\rightarrow 0.65$}
\includegraphics[height=.18\textheight]{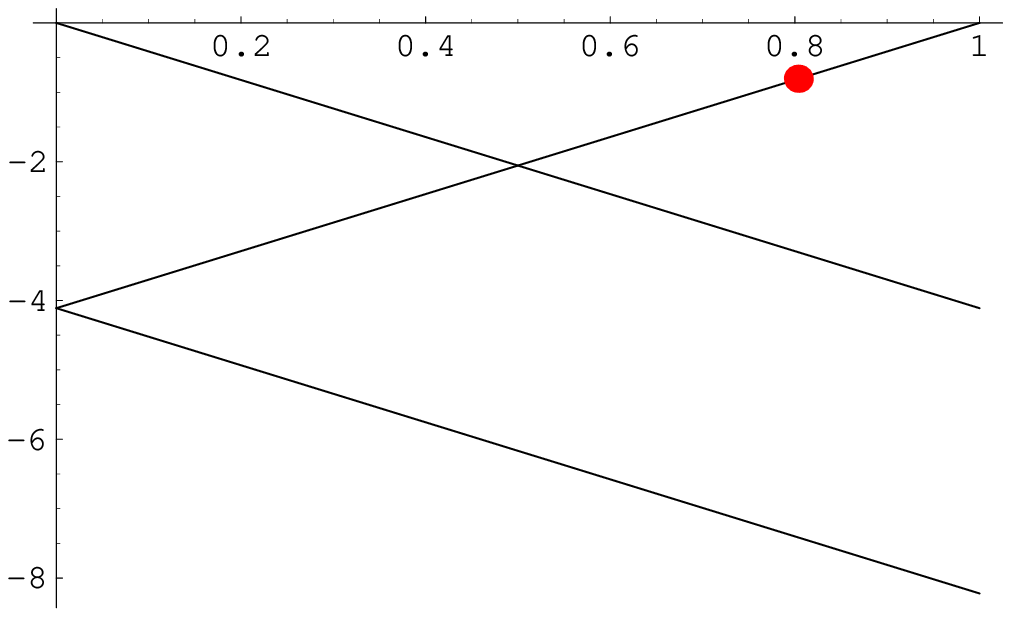}
\hspace{-0.5cm} $a_p\rightarrow 0.65$ \caption{ The boundary in
the $a_p,a_n$ plane extracted from the data for the target
$^{127}I$ is shown assuming  that the amplitudes are relatively
real.  The scale depends on the event rate and the LSP  mass.
Shown here is the scale for $m_{\chi}=100$ GeV.  Note that the
allowed region is confined when the amplitudes are of the same
sign (left plot), but they are not confined when the amplitudes
are of opposite sign. The allowed space now is i) The small
triangle and ii) The space between the two parallel lines and on
the right of the line that intercepts them. We also indicate by a
dot the point $a_p=-a_n$ favored by the spin structure of the
nucleon. The nuclear ME employed were those of Ressel and Dean
(see table \ref{table.spin})
 \label{gpgnamp} }
\end{center}
\end{figure}

\begin{figure}
\begin{center}
\rotatebox{90}{\hspace{0.0cm} $\sigma_n \rightarrow 5.0\times
10^{-3}$ pb}
\includegraphics[height=.15\textheight]{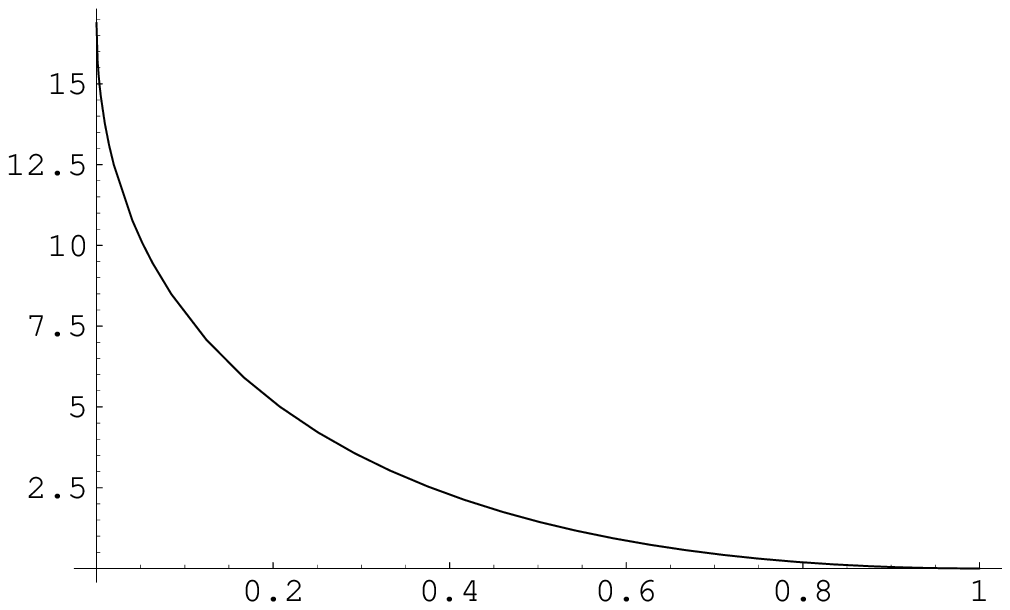}
\hspace{-0.5cm} $\sigma_p\rightarrow 5.0\times 10^{-3}$ pb
\rotatebox{90}{\hspace{0.0cm} $\sigma_n\rightarrow 5.0\times
10^{-3}$ pb }
\includegraphics[height=.15\textheight]{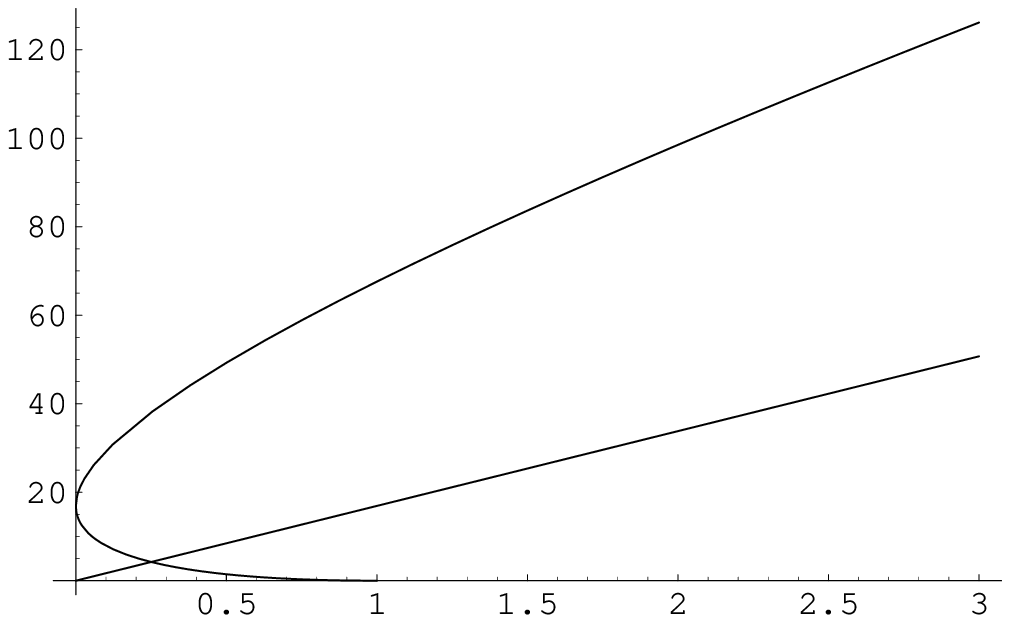}
\hspace{-0.5cm} $\sigma_p\rightarrow 5.0\times 10^{-3}$ pb
\caption{ The same as in Fig. \ref{gpgnamp}  for the
$\sigma_p,\sigma_n$ plane.
 On the left the allowed region  is that below the curve (the amplitudes are relatively real  and
 have the same sign) .  In the plots on the right the amplitudes are relatively real and of opposite sign. The allowed region is i) between the higher segment of the hyperbola and the straight line and ii) Between the straight line and the lower segment of the curve.
%Thus in the lower "triangle", appearing both on the left and the right figures, the amplitudes are %unambiguously of the same sign. In the other triangle the sign cannot be inferred. In the rest of the space
%one can infer that the amplitudes  have the opposite sign.
% The point favored by the structure of the\ nucleon is also indicted.
 The
nuclear ME employed were those of Ressel and Dean (see table
\ref{table.spin})
 \label{gpgnsigma} }
\end{center}
\end{figure}

\begin{figure}
\begin{center}
\rotatebox{90}{\hspace{0.0cm} $\sigma_n\rightarrow 5.0\times
10^{-3}$ pb}
\includegraphics[height=.3\textheight]{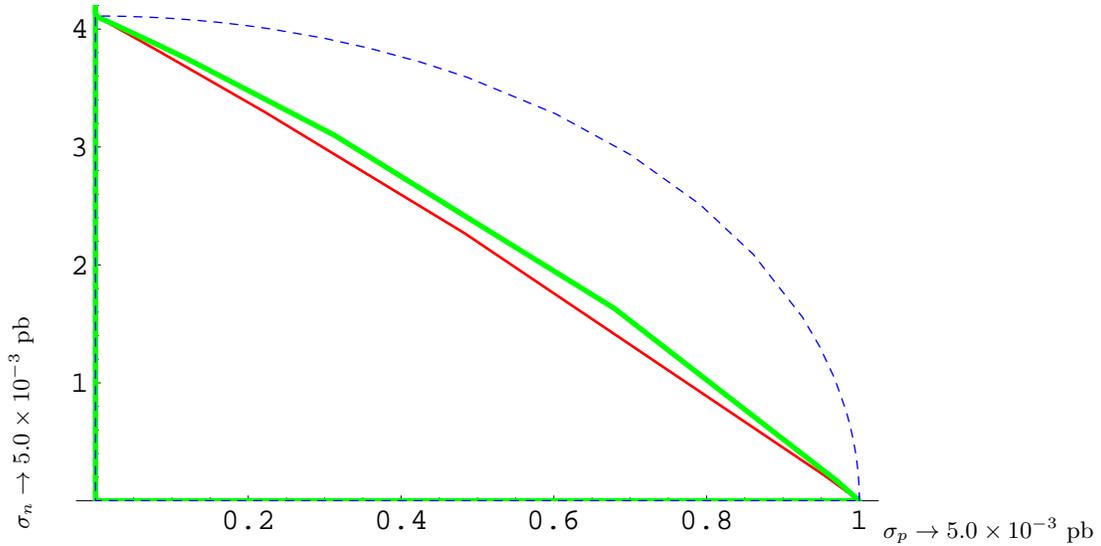}
\hspace{-0.5cm} $\sigma_p\rightarrow 5.0\times 10^{-3}$ pb
\caption{ The same as in Fig. \ref{gpgnsigma}  assuming that the
amplitudes are not relatively real, but are characterized by a
phase difference $\delta$. The allowed space is now confined. The
results shown for the thin solid, thick solid and dashed curves
correspond  to $\delta=\pi /3,~\pi /6$ and $\pi /2$ respectively .
 \label{sigmadelta} }
\end{center}
\end{figure}

The situation is exhibited in  Figs \ref{gpgnamp}-\ref{sigmadelta}
in the interesting case of the A=127 system using the nuclear
matrix elements of Ressel {\it et al} given in Table
\ref{table.spin}. One can understand the asymmetry in the plot due
to the fact that $\Omega_p$ is much larger than $\Omega_n$. In
other words if $\sigma_p$ happens to be  very small a large
$\sigma_n$ will be required to accommodate the data.  In the
example considered here, however, the extreme values differ only
by $20\%$
 from the values on the axes,
which arise if one assumes that one mechanism at a time (proton or neutron) dominates.
\subsection{ Spin matrix elements of opposite sign and $|\Omega_p|\gg |\Omega_n|$}
\begin{figure}
\begin{center}
\rotatebox{90}{\hspace{0.0cm} $a_n\rightarrow 0.3$}
\includegraphics[height=.18\textheight]{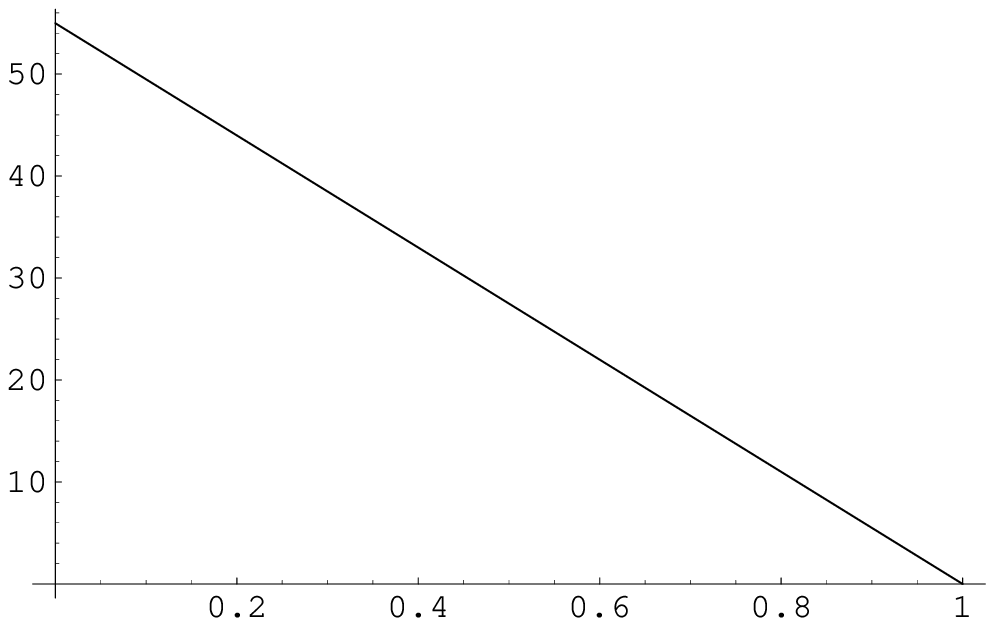}
\hspace{-0.5cm} $a_p\rightarrow 0.3$
\rotatebox{90}{\hspace{0.0cm} $a_n\rightarrow 0.3$}
\includegraphics[height=.18\textheight]{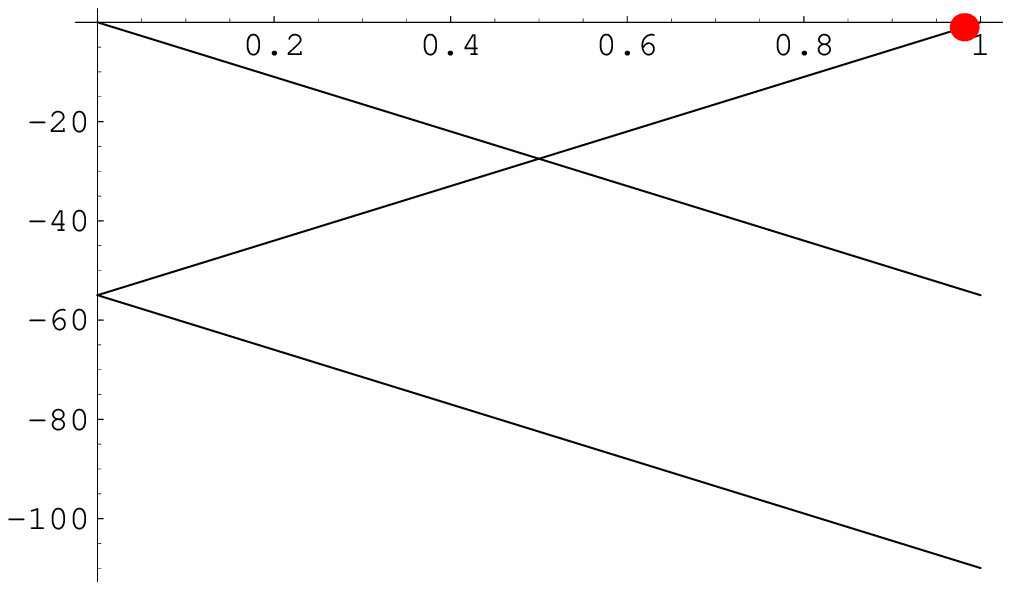}
\hspace{-0.5cm} $a_p\rightarrow 0.3$
\caption{ The same as in
Fig. \ref{gpgnamp} for the target $^{19}F$. The matrix elements
employed are those of table \ref{table.spin}. Since now the proton
and neutron spin ME have opposite signs, the conclusions about the
relative sign of the elementary amplitudes $a_p,a_n$ are opposite
of those arrived in  Fig.  \ref{gpgnamp}. The limit extracted on
the neutron amplitude is poor, since the neutron spin ME is tiny.
 \label{gpgnamp19} }
\end{center}
\end{figure}

\begin{figure}
\begin{center}
\rotatebox{90}{\hspace{0.0cm} $\sigma_n\rightarrow  2.3\times
10^{-3}$ pb}
\includegraphics[height=.15\textheight]{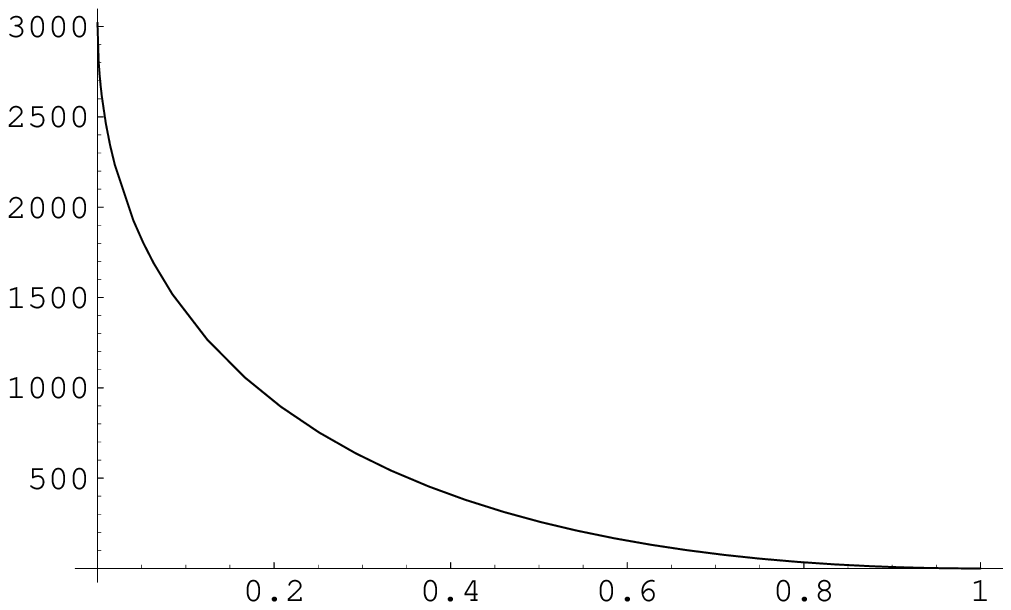}
\hspace{-0.5cm} $\sigma_p\rightarrow 2.3\times 10^{-3}$ pb
\rotatebox{90}{\hspace{0.0cm} $\sigma_n\rightarrow 2.3\times
10^{-3}$ pb}
\includegraphics[height=.15\textheight]{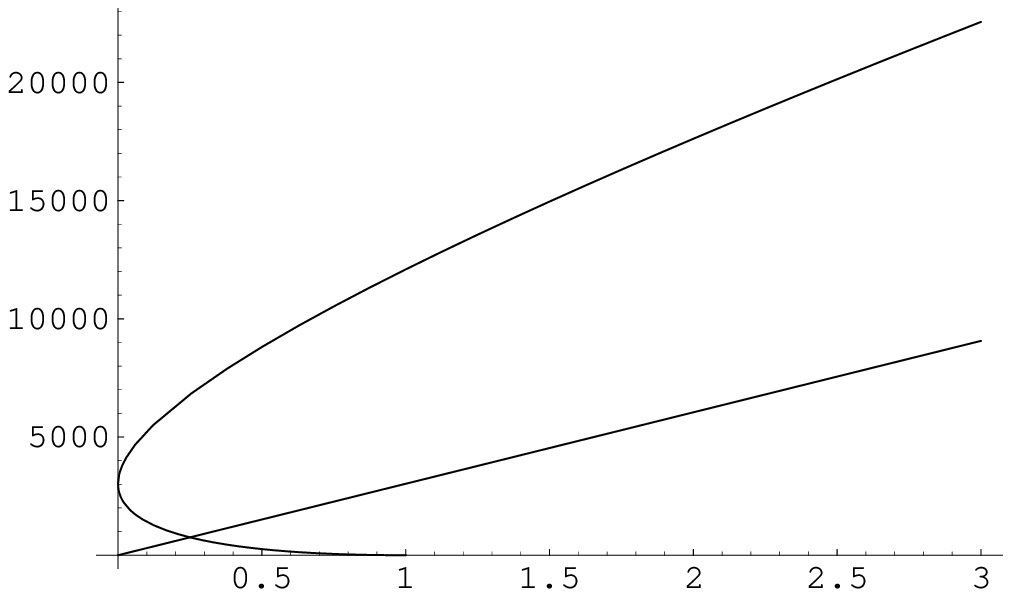}
\hspace{-0.5cm} $\sigma_p\rightarrow 2.3\times 10^{-3}$ pb
\caption{ The same as in Fig. \ref{gpgnsigma} for $^{19}F$ with
remarks about the signs of the elementary amplitudes as in Fig.
\ref{gpgnamp19}. The matrix elements employed are those of table
\ref{table.spin}.
 \label{gpgnsigma19} }
\end{center}
\end{figure}

\begin{figure}
\begin{center}
\rotatebox{90}{\hspace{0.0cm} $\sigma_n\rightarrow 2.3\times
10^{-3}$ pb}
\includegraphics[height=.3\textheight]{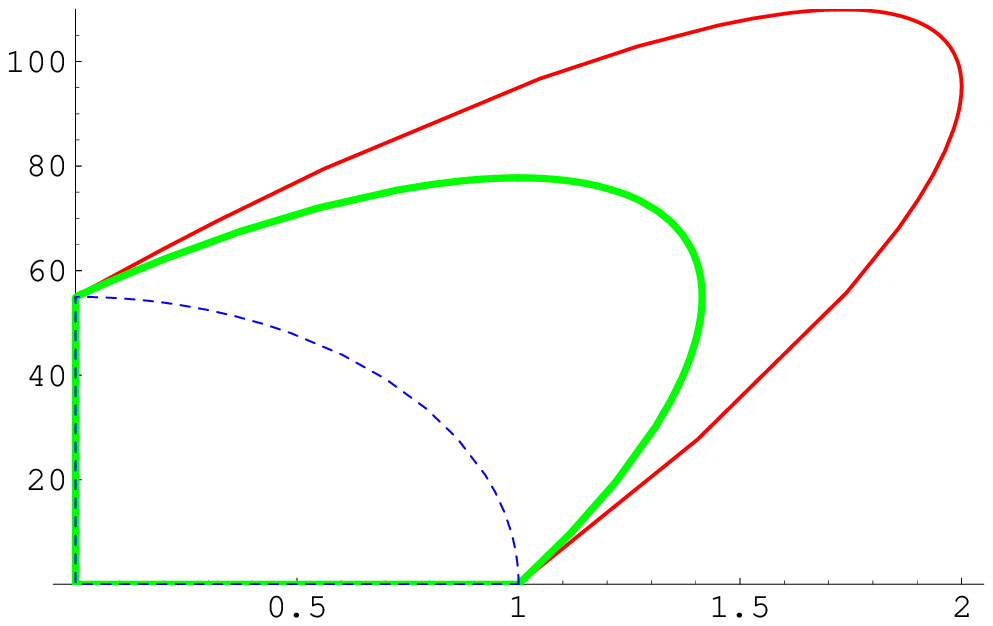}
\hspace{-0.5cm} $\sigma_p\rightarrow 2.3\times 10^{-3}$ pb
\caption{ The same as in Fig. \ref{sigmadelta}  for $^{19}F$ with
remarks about the signs of the elementary amplitudes as in Fig.
\ref{gpgnamp19}. Note that now that the spin matrix elements are
of opposite sign the extracted limits on the nucleon cross section
are factor of two larger than assuming only one amplitude.
 \label{sigmadelta19} }
\end{center}
\end{figure}

This situation occurs in the case of the target $^{19}F$. As we have already mentioned the corresponding spin matrix elements shown in \ref{table.spin} are quite reliable. The obtained results are shown in Figs
\ref{gpgnamp19}-\ref{sigmadelta19}
\subsection{ Spin matrix elements of the same sign and $\Omega_n\gg \Omega_p$}
\begin{figure}
\begin{center}
\rotatebox{90}{\hspace{0.0cm} $a_n\rightarrow 200$}
\includegraphics[height=.18\textheight]{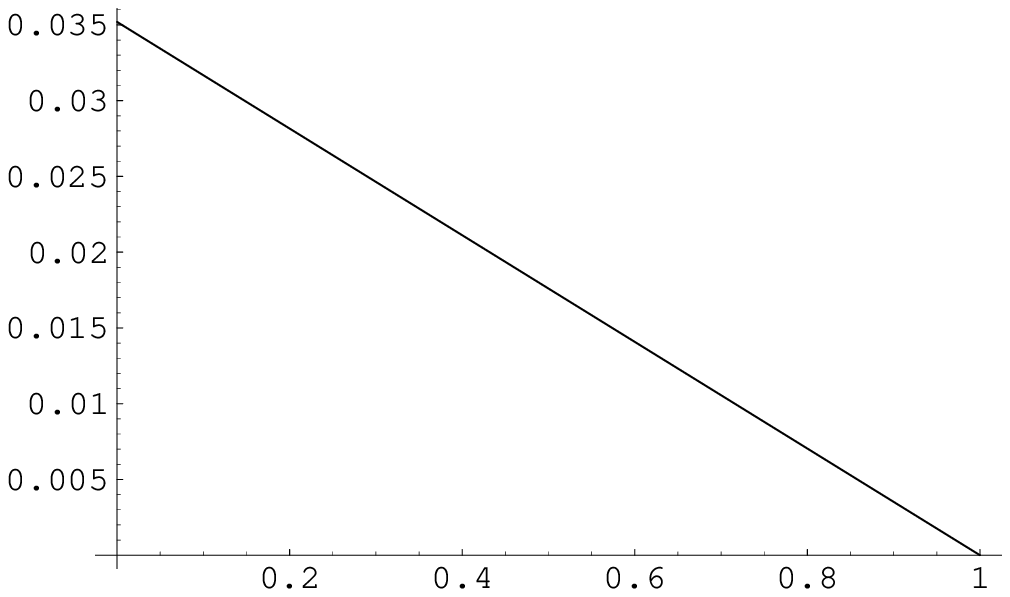}
\hspace{-0.5cm} $a_p\rightarrow 200$
 \rotatebox{90}{\hspace{0.0cm}
$a_n\rightarrow 200$}
\includegraphics[height=.18\textheight]{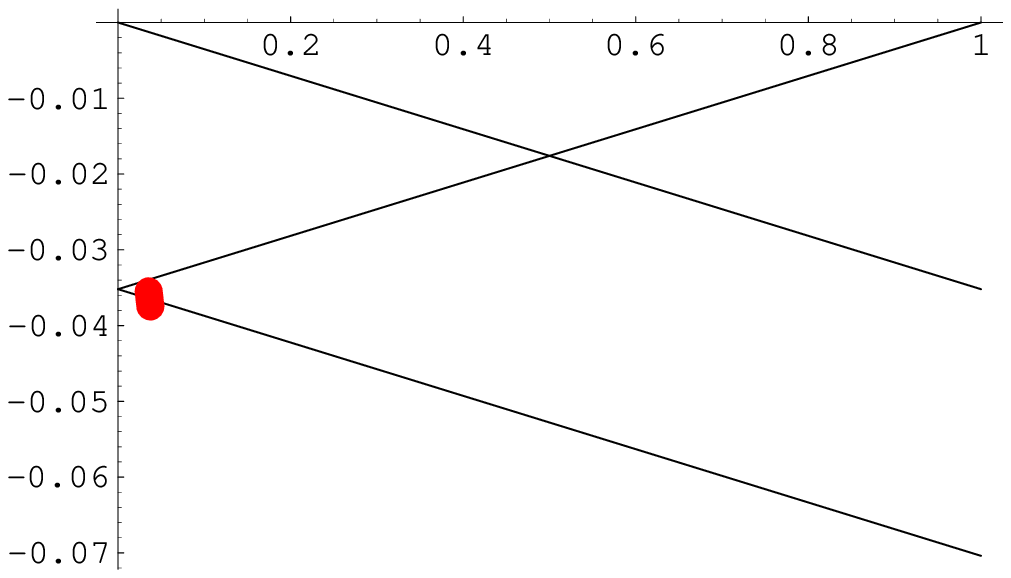}
\hspace{-0.5cm} $a_p\rightarrow 200$ \caption{ The same as in Fig.
\ref{gpgnamp} for the target $^{73}Ge$. The matrix elements
employed are those of table \ref{table.spin}. The proton and
neutron spin ME have the same sign, but now $\Omega_n\gg
\Omega_p$. Thus the limit on $a_n$ is more stringent than that on
$a_p$.
 \label{gpgnamp73} }
\end{center}
\end{figure}

\begin{figure}
\begin{center}
\rotatebox{90}{\hspace{0.0cm} $\sigma_n\rightarrow 1.6$ pb}
\includegraphics[height=.15\textheight]{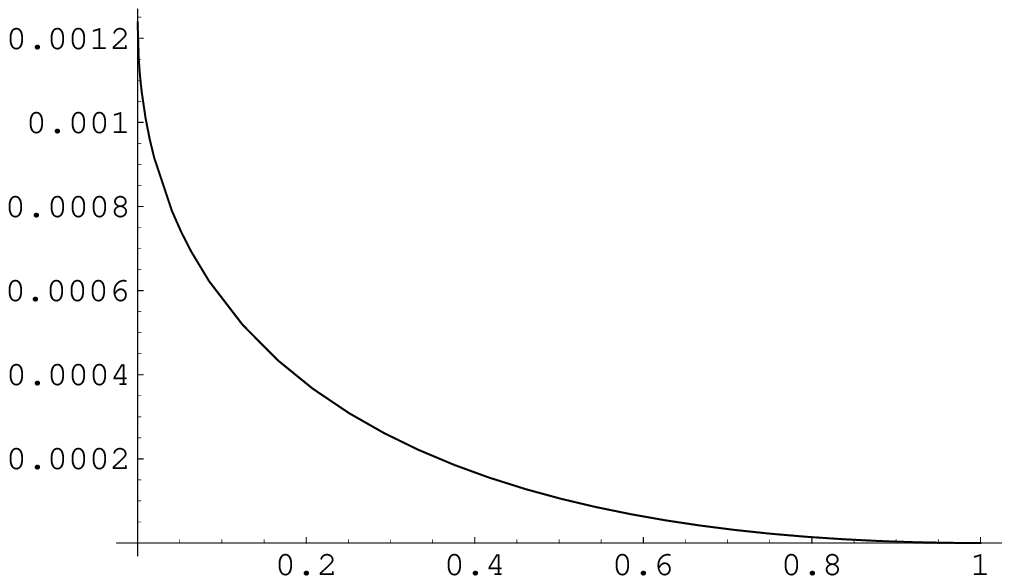}
\hspace{-0.5cm} $\sigma_p\rightarrow 1.6$ pb
\rotatebox{90}{\hspace{0.0cm} $\sigma_n\rightarrow 1.6$ pb }
\includegraphics[height=.15\textheight]{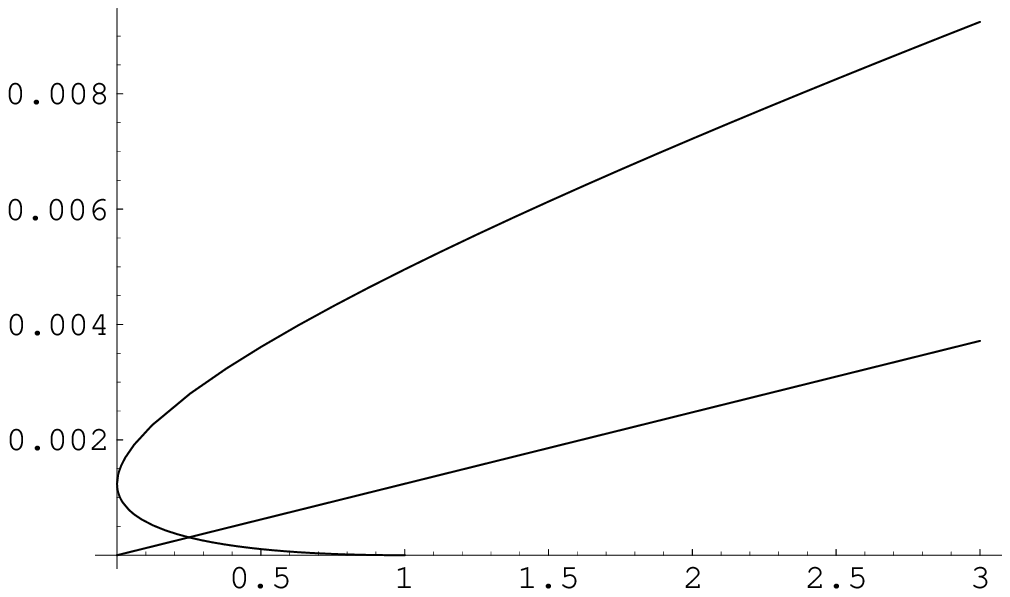}
\hspace{-0.5cm} $\sigma_p\rightarrow 1.6$ pb
\caption{ The same as in Fig. \ref{gpgnsigma} for $^{73}Ge$ with
similar remarks  about the nucleon cross sections  as in Fig.
\ref{gpgnamp73} for the amplitudes. The matrix elements employed
are those of table \ref{table.spin}.
 \label{gpgnsigma73} }
\end{center}
\end{figure}

\begin{figure}
\begin{center}
\rotatebox{90}{\hspace{0.0cm} $\sigma_n\rightarrow 1.6$ pb}
\includegraphics[height=.35\textheight]{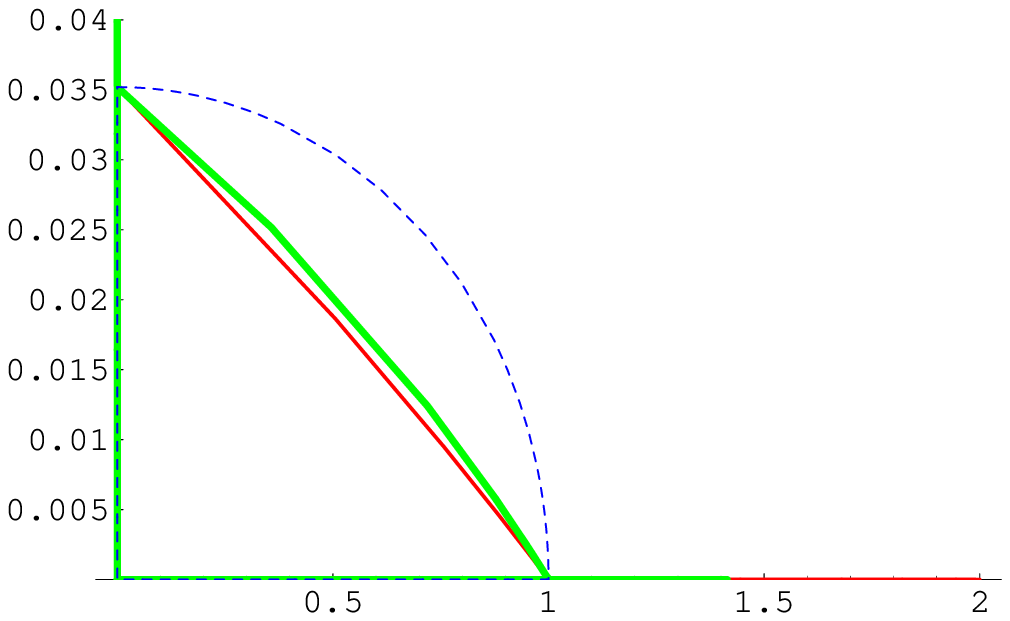}
\hspace{-0.5cm} $\sigma_p\rightarrow 1.6$ pb
\caption{ The same as in Fig. \ref{sigmadelta}  for $^{73}Ge$ with
remarks about the nucleon cross sections signs as in Fig.
\ref{gpgnsigma73}.
 \label{sigmadelta73} }
\end{center}
\end{figure}

This situation occurs in the case of the target $^{73}Ge$ (see table \ref{table.spin}). The obtained results are shown in Figs
\ref{gpgnamp73}-\ref{sigmadelta73}. Now the most stringent limit is imposed on the neutron cross section.
\subsection{ Spin matrix elements of opposite sign and $|\Omega_n|\gg |\Omega_p|$}
\begin{figure}
\begin{center}
%\rotatebox{90}{\hspace{0.0cm} $a_n\rightarrow$}
\includegraphics[height=0.4\textheight]{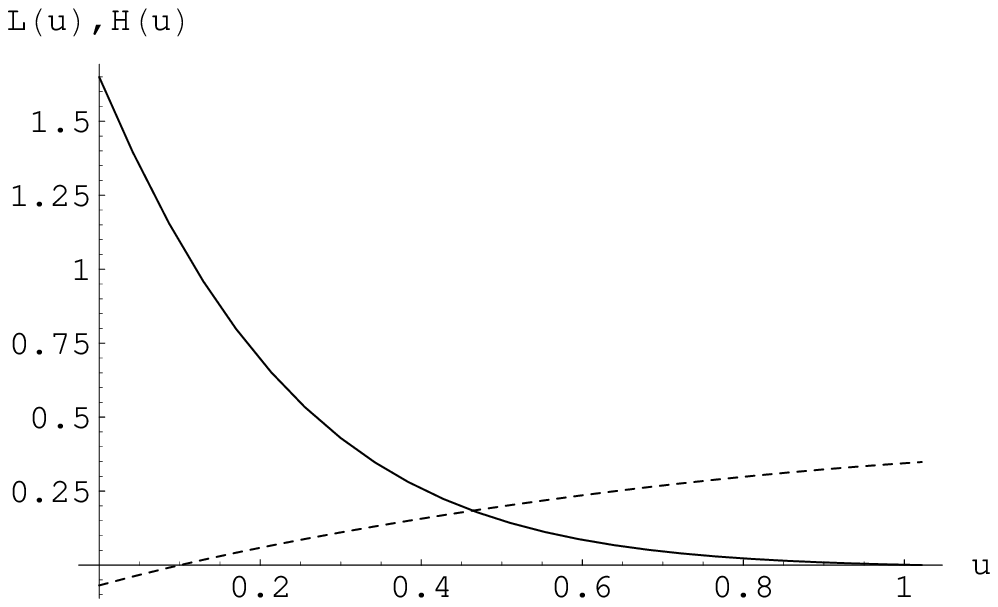}
\hspace{-0.5cm} $Q\rightarrow 26$ keV.
%\rotatebox{90}{\hspace{0.0cm} $a_n\rightarrow$}
%\includegraphics[height=.2\textheight]{ampboundary3.eps}
%\hspace{-0.5cm} $a_p\rightarrow$
\caption{The quantities  $L(u)$ (continuous curve) and $H(u)$  (dashed curve), as functions of the dimensionless energy transfer $u$, $Q=Q_0 u$, $Q_0=26$ keV. For their definition see Eq. \ref{difmodulation}.
 \label{fig:difmod} }
\end{center}
\end{figure}

\begin{figure}
\begin{center}
\rotatebox{90}{\hspace{0.0cm} $a_n\rightarrow 88$ }
\includegraphics[height=.18\textheight]{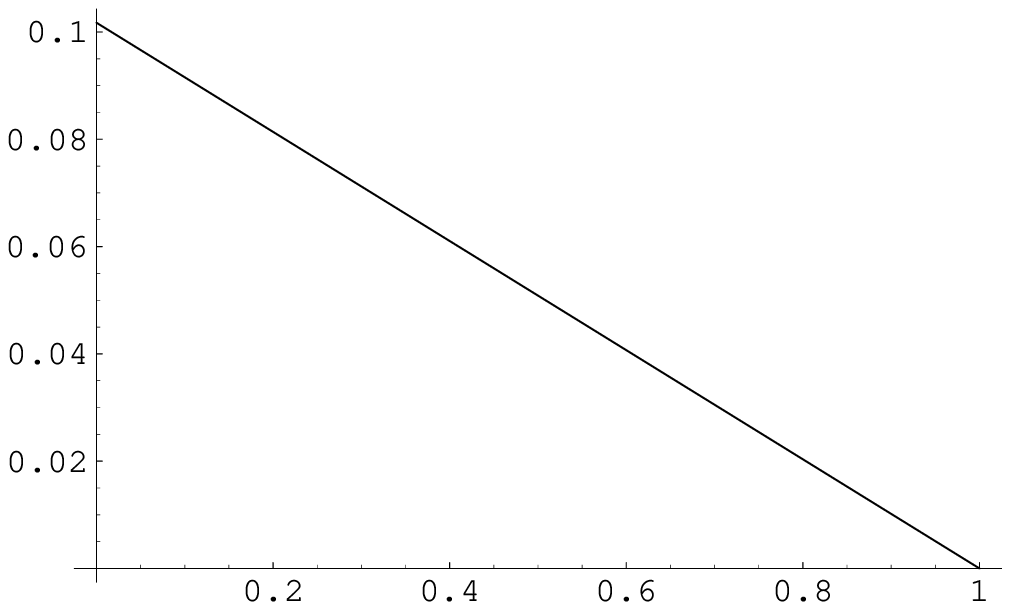}
\hspace{-0.5cm} $a_p\rightarrow 88$ \rotatebox{90}{\hspace{0.0cm}
$a_n\rightarrow 88$}
\includegraphics[height=.18\textheight]{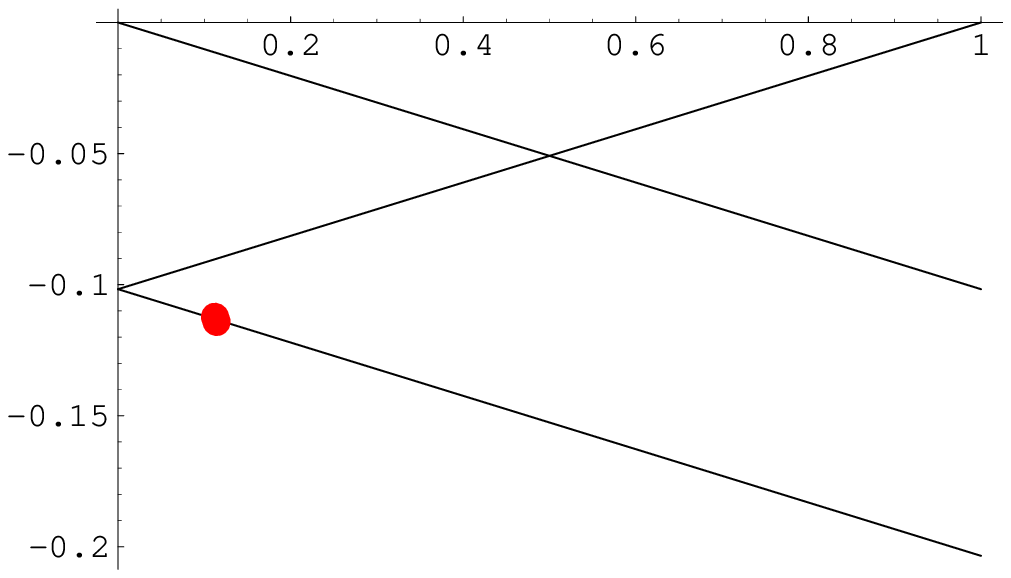}
\hspace{-0.5cm} $a_p\rightarrow 88$ \caption{ The same as in Fig.
\ref{gpgnamp73} for the target $^{3}He$, except that the spin
matrix elements have opposite sign (now $|\Omega_n|>>\Omega_p$).
As a result the inferred relative sign of the amplitudes is
opposite to those of Fig.\ref{gpgnamp73}.  The matrix elements
employed are those of table \ref{table.spin}.
 \label{gpgnamp3} }
\end{center}
\end{figure}

\begin{figure}
\begin{center}
\rotatebox{90}{\hspace{0.0cm} $\sigma_n\rightarrow 0.68$ pb}
\includegraphics[height=.15\textheight]{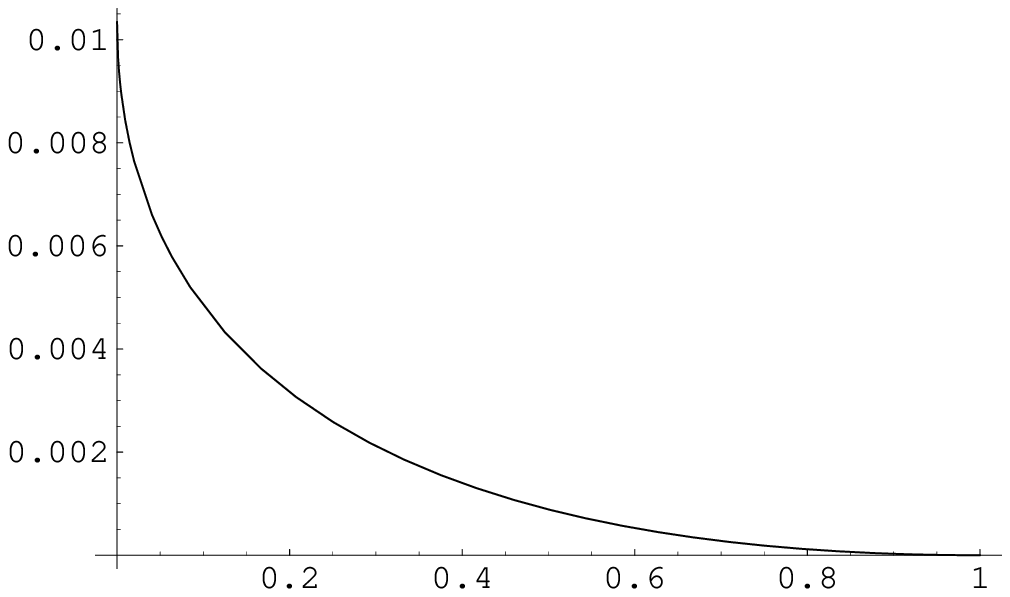}
\hspace{-0.5cm} $\sigma_p\rightarrow 0.68$ pb
\rotatebox{90}{\hspace{0.0cm} $\sigma_n\rightarrow 0.68$ pb }
\includegraphics[height=.15\textheight]{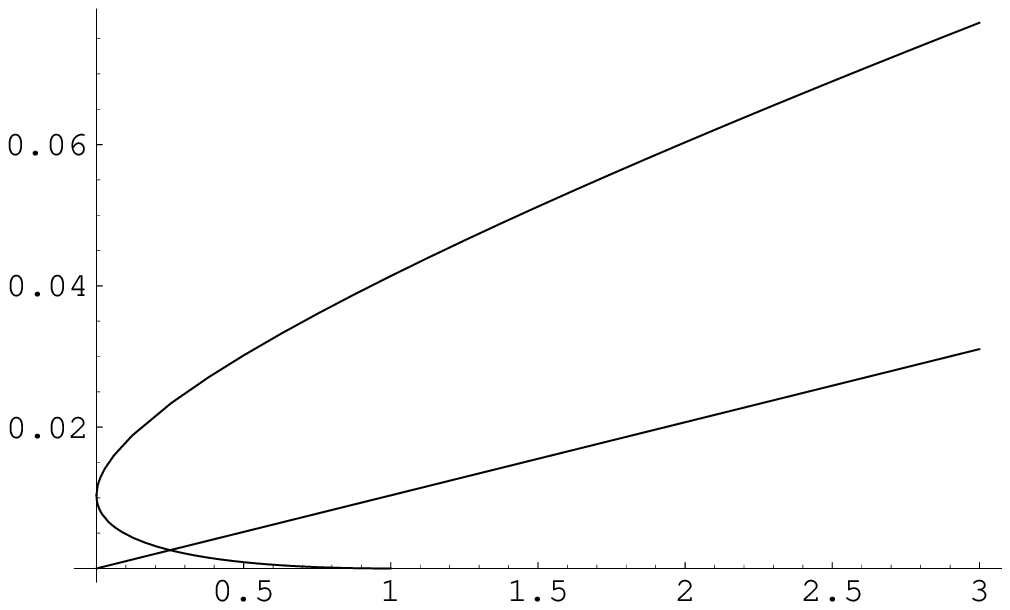}
\hspace{-0.5cm} $\sigma_p\rightarrow 0.68$ pb
\caption{ The same as in Fig. \ref{gpgnsigma73} for $^{3}He$ with
the same remarks about the nucleon cross sections as in Fig.
\ref{gpgnamp73}. The matrix elements employed are those of table
\ref{table.spin}.
 \label{gpgnsigma3} }
\end{center}
\end{figure}

\begin{figure}
\begin{center}
\rotatebox{90}{\hspace{0.0cm} $\sigma_n\rightarrow 0.68$ pb}
\includegraphics[height=.25\textheight]{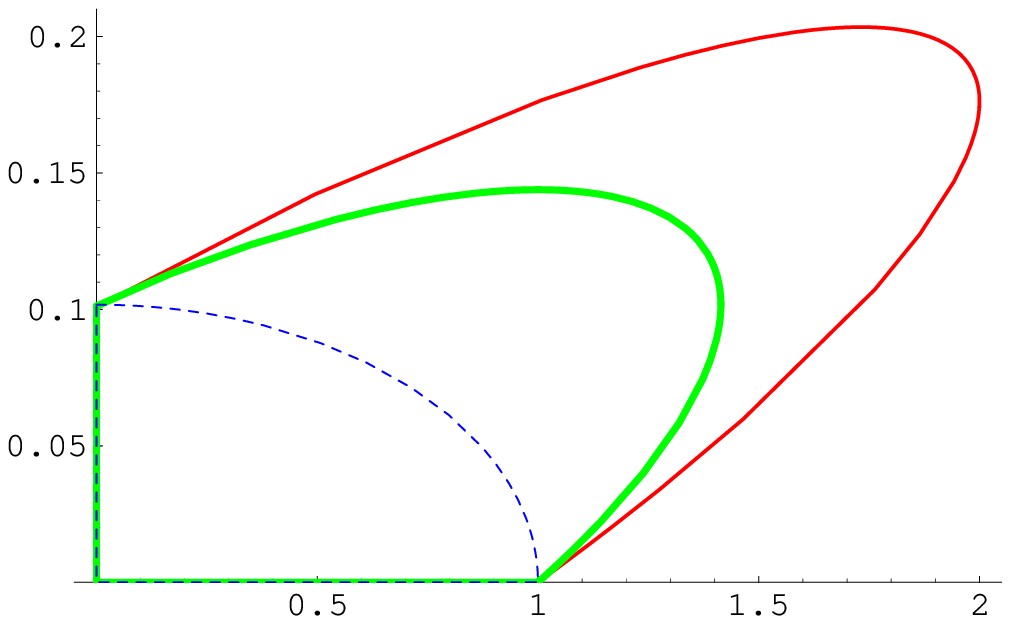}
\hspace{-0.5cm} $\sigma_p\rightarrow 0.68$ pb
\caption{ The same as in Fig. \ref{sigmadelta}  for $^{3}He$ with
remarks about the signs
 of the elementary amplitudes as in Fig. \ref{gpgnsigma3}. Note that the extracted
 limits are a factor of two larger than assuming only one amplitude.
 \label{sigmadelta3} }
\end{center}
\end{figure}

This situation occurs in the case of the target $^{3}He$. The quantities $t$ and $h$ are  essentially independent of the LSP mass, since the LSP is expected to be much heavier than the nuclear mass. We find
$t=2.33$ and $h=0.045$ as the modulation amplitude. The latter is a bit larger than in heavier nuclei.  This means that $c_{spin}(3,\mu_r(3))=3 t=7.00$. Likewise the parameters $\sigma _{bound}(3)~,~r(m_{\chi},3)$ are also constants, $\sigma _{bound}(3)=$, $r(m_{\chi},3)=1$. Thus the expected rate is expected a bit smaller. This nucleus, however, has definite experimental advantages \cite{Santos}. Furthermore the nuclear matrix elements can be calculated very reliably. The differential tate is also independent of the LSP mass. Thus it can simply be exhibited  in the form:
\beq
\frac{dR}{du}=R L(u) \left[1+H(u) \cos{\alpha}\right]
\label{difmodulation}
\eeq
Where $L(u)$, integrated from $0$ to $u_{max}=1$ yields unity, and $H(u)$ is the ratio of the modulated by the unmodulated differential rate. The situation is shown in Fig. \ref{fig:difmod}.
The  constraints on the elementary amplitudes and the nucleon spin cross sections are shown in Figs
\ref{gpgnamp3}-\ref{sigmadelta3}. Again the most stringent limit is imposed on the neutron cross section.
\section{Conclusions}
 In the present paper we have studied the contribution of the axial current to the direct detection of the SUSY dark matter focusing on the popular targets $^{127}$I, $^{73}$Ge, $^{19}$F and $^3$He. The nuclear structure of these targets seems to favor one component, either neutron, $|\Omega_n| \gg |\Omega_p|$,   or proton, $|\Omega_p| \gg |\Omega_n|$ (see table \ref{table.spin}) . The real question is the size and the relative importance of the elementary amplitudes $a_p$ and $a_n$. These result from the combination of two factors:
 \begin{itemize}
 \item The relevant amplitudes at the quark level.\\
 In the case Z-exchange the  the isoscalar amplitude is zero. In the case of the s-quark exchange the relative importance of the two amplitudes depends on the details of the SUSY parameter space. A priori there is no reason to prefer one amplitude over the other.
 \item Going from the quark to the nucleon level. \\
The  structure of the nucleon tends to favor the isovector component.  Thus, barring unusual circumstances  at the quark level, the amplitudes $a_p$ and $a_n$ have opposite sign. 
 Unfortunately none of the nuclear systems considered satisfies the most favorable condition $\Omega_n=-\Omega_p$. Nonetheless in this case one will be able to simply extract the amplitudes $a_p$ and $a_n$ from the experimental data, if and when they become available. 
\end{itemize}
  In the most general case the extraction of the elementary amplitudes from the data is not going to be trivial. One can only make  exclusion plots in the two dimensional ($a_p$, $a_n$)  and 
($\sigma_p$, $\sigma_p$) planes. By comparing such plots involving various targets one may be able to extract both amplitudes. In this work we have drawn such plots,  using as input the nuclear spin response functions ( static spin ME and the relevant form factors obtained in the context of the shell model) in conjunction with the experimental limit, taken to be $\prec 16$ events per kg-y for all targets considered. 

\section*{Acknowledgements}
This work was supported by European Union under
the contract MRTN-CT-2004-503369 as well as the program PYTHAGORAS-1. The latter is part of the
Operational Program for Education and Initial Vocational Training of the
Hellenic Ministry of Education under the 3rd Community Support Framework
and the European Social Fund. The author is indebted to professor Miroslav Finger for his hospitality during the Praha-spin-2005 conference.
 
%^\bibliographystyle{myprsty}
%\bibliography{TeX}
%printindex

\end{document}